\documentclass[reprint,
 amsmath,amssymb,
 aps, physrev,
]{revtex4-2}
\usepackage{multirow}
\usepackage{graphicx}
\usepackage{dcolumn}
\usepackage{bm}
\usepackage{soul}
\usepackage{subcaption}
\usepackage{xcolor}
\usepackage{hyperref}

\newcommand{\RNum}[1]{\uppercase\expandafter{\romannumeral #1\relax}}

\begin{document}

\preprint{prb}

\title{Particle Dynamics in Constant Synthetic Non-Abelian Fields}

\author{Subramanya Bhat K. N.}
\altaffiliation[]{}
\email{subramanyabhatkn@gmail.com}
\author{Amita Das}%
\email{amita@iitd.ac.in}
\affiliation{%
 Indian Institute of Technology Delhi, New Delhi, India 
}%

\author{V Ravishankar}
\email{v.ravishankar@iiitdwd.ac.in}
\affiliation{Indian Institute of Information Technology Dharwad, Dharwad, India}

\author{Bhooshan Paradkar}
\email{bhooshan.paradkar@cbs.ac.in}
\affiliation{%
  UM-DAE Centre for Excellence in Basic Sciences, University of Mumbai 
}%
%


\date{\today}

\begin{abstract}
Yang–Mills theory has extended well beyond its original role in describing the strong force and now emerges as an effective theory in condensed matter, ultracold atomic, and photonic systems. In these systems, the theory has been successful in explaining phenomena such as the spin-Hall effect, spin transport, and controlling the polarisation of light. Moreover, the ability to engineer and control synthetic non-Abelian gauge fields in these systems enables us to explore aspects of gauge dynamics inaccessible to high-energy experiments. In all the above mentioned cases, the state of the system evolves in an effective external Yang-Mills field. Thus, the study of test particle dynamics in such background fields is interesting in both the classical and quantum mechanical regimes. The background non-Abelian (color) gauge fields considered in this study are constant, and they generate uniform color magnetic fields or combined color electric and magnetic fields- which are relevant configurations. Despite the apparent simplicity of these backgrounds, the coupled evolution of real space motion and internal color degrees of freedom results in rich, nontrivial behaviour that is qualitatively distinct from the electrodynamic (Abelian) case, such as unbounded trajectories in a constant color magnetic field. In particular, particle trajectories encode signatures of the underlying gauge sources. Finally, the classical dynamics presented in this paper serves as a precursor to the complete quantum mechanical treatment to follow.

\end{abstract}

\maketitle


\section{\label{sec:level1}Introduction} 
Yang–Mills (YM) theory, originally introduced as a non-Abelian generalisation of Maxwell’s electrodynamics (ED) to describe the strong interaction \cite{YangMills-original}, has since emerged as a unifying framework across a wide range of physical systems. Beyond its foundational role in high-energy physics, it is now well established that effective non-Abelian gauge fields arise naturally in condensed-matter, atomic, photonic and metamaterial systems \cite{ColdAtomPRLOsterloh,ColloquiumArtificialG,YiYangSyntAndObsv,PolimenoExptNAinPerovskite,YiYangNAinLightandSound,YMSpinTAN}. In these systems, the gauge fields are emergent. They encode how internal degrees of freedom, such as spin (in condensed matter system), pseudospin (in laser-atom interactions) and polarisation or orbital angular momentum (in optical systems), couple to the particle's motion. This realisation has allowed the control of the internal degrees of freedom. This not only has technological applications\cite{PhysRevApplied.SHEswitch,SHE.switch,PSHE.PQE}, but also allows us to explore those features of the gauge dynamics which are not easily accessible to high-energy physics experiments.

In condensed matter systems, effective non-Abelian potentials arise in several situations. The non-Abelian theory has been employed to describe many phenomena such as superconductivity \cite{chosuper}, spin transport, and the spin-Hall effect\cite{YMSpinTAN}. More prominently, in systems with spin-orbit coupling, one can realise constant non-Abelian gauge fields. Here, the electron spin plays the role of the non-Abelian charge (color degree of freedom), and the emergent gauge symmetry can be $SU(2)$ or $U(2)$. The electron responds to the resultant color electric and magnetic fields.  A well-known example is the Rashba type of spin-orbit coupling. This interaction is realised experimentally in many materials \cite{Nitta1997,grundler2000large,caviglia2010tunable}. The strength of the gauge potentials can vary over a large range of values.

A similar generation of synthetic non-Abelian gauge fields is possible in ultracold atoms. These systems provide good control over the generation and control of synthetic gauge potentials. Importantly, these systems allow the realisation of specific non-Abelian field configurations that are directly relevant to the present study. Here, the role of color charge is taken by the internal energy levels of the atom. By appropriately choosing the system, type of laser coupling, or other external parameters, such as magnetic fields, one can realise a variety of gauge symmetries such as $SU(2), U(2), SO(3)$ or $SU(3)$\cite{PRL.GenMag,ColdAtomPRLOsterloh,ColloquiumArtificialG,rico2018so,anderson2012synthetic,anderson2013magnetically,madasu2025experimental}. In these systems, people have proposed the possible observation of non-Abelian Aharonov-Bohm effect \cite{ColdAtomPRLOsterloh}, and it has been demonstrated experimentally recently\cite{liang2024chiral}. 

Synthetic non-Abelian gauge potentials have also been realised in photonic and metamaterial systems, where polarisation (spin or orbital-angular momentum of light) constitutes the internal degrees of freedom. In these systems, different non-Abelian gauge configurations can be realised by tweaking the permittivity and permeability tensors in the presence of an electromagnetic wave. Different configurations have been proposed theoretically \cite{liu2015gauge,chen2019non}, of which some configurations have been realised experimentally \cite{YiYangSyntAndObsv,liu2025general}. These systems are also of particular interest to this study as they host $SU(2)$ gauge fields and allow multiple configurations involving multiple combinations of color electric and color magnetic fields. 

From a theoretical standpoint, the first step would be to explore the dynamics of a test particle in the presence of color fields. While a full quantum mechanical treatment provides the most accurate description, the classical description is itself instructive and merits independent study. Even at the classical level, the coupled evolution of a particle’s real-space motion and its internal (color) degrees of freedom results in dynamics which is qualitatively distinct from the Abelian (ED) case, even when the color fields are homogeneous. In particular, it serves as a precursor and an instructive guide to understand the results of the complete quantum mechanical treatment.

In a complementary case, we have studied the dynamics of a test particle in the presence of a constant color electric field generated by spatially uniform YM gauge potentials \cite{KN_2025}. This study was conducted with the main focus on the quark-gluon plasma produced in heavy-ion collisions. That apart, our studies showed that, despite the absence of spatial or temporal dependence of the color electric field, the particle's trajectory can be bounded, a behaviour which is opposite to what would happen in ED. It was further shown that the trajectory of the particle can, in fact, be used to characterise the underlying gauge structure.

In the same spirit, our present work extends the previous work to the test particle dynamics in the presence of constant non-Abelian gauge potentials, which produce (a) only a color magnetic field, or (b) a combination of both color electric and color magnetic fields. As hinted earlier, these configurations may be realised experimentally, some examples being \cite{Nitta1997,YiYangSyntAndObsv,chen2019non,liu2025general}. Our study mainly focuses on the $SU(2)$ gauge symmetry. Since the study is classical, the relevant gauge symmetry becomes $SO(3)$. 

The paper is organised as follows. In section.~\ref{sec:classical-YM}, we introduce the class of non-Abelian fields considered in this work. Sections~\ref{sec:Magdyn-1D} and~\ref{sec:Magdyn-3D} are devoted to particle dynamics in maximally non-Abelian magnetic fields. Section~\ref{sec:Combdyn} discusses motion in combined non-Abelian electric and magnetic backgrounds. We conclude with a discussion in section~\ref{sec:conclusions}.

\section{\label{sec:classical-YM}Maximally Non-Abelian Fields}
Following \cite{KNmine}, we first set up the notations employed in this paper\footnote{For the background review of Yang-Mills dynamics, one can refer to the following standard treatments \cite{WongClassicalYM-Isospin,Boozer,YMSpinTAN,KNmine}}. We employ the standard vectorial notation for all operations in the internal space. Since the gauge group is $SO(3)$, we take the associated internal space  $\mathbb{R}^3$ to be spanned by an orthonormal basis $\{\hat{e}_1, \hat{e}_2 ,\hat{e}_3\}$. The Latin indices are used to represent the spatial components. Accordingly, the color charge gets denoted by $\vec{q}$, the color vector potential by $\vec{A}_i$ and the color scalar potential by $\vec{\phi}$. The color electric $(\vec{E}_i)$ and color magnetic fields ($\vec{B}_i$) are then expressed as
\begin{subequations}
    \label{eq:fdef}
    \begin{eqnarray}
        \vec{E} _i &=& -\partial_i \vec{\phi} - \frac{1}{c}\partial_0 \vec{A}_i + g\, \vec{A}_i \times \vec{\phi} \label{eq:edef} \\
        \vec{B}_i &=& \partial_j \vec{A}_k - \partial_k \vec{A}_j - g\, \vec{A}_j \times \vec{A}_k \label{eq:bdef}
    \end{eqnarray}
\end{subequations}
where $g$ is the non-Abelian coupling constant. In the presence of these fields, the dynamics of a test particle with mass $m$ and charge $\vec{q}$ is described by a set of Lorentz and Wong equations \cite{WongClassicalYM-Isospin}, given by
\begin{subequations}
    \label{eq:dyn}
    \begin{eqnarray}
    \frac{d P_i}{dt} &=& g\vec{q} \cdot \left(\vec{E}_i + \epsilon_{ijk}\frac{v_j}{c}\vec{B}_k \right), \label{eq:lz} \\
    \frac{d\vec{q}}{dt} &=& gc \vec{q} \times \left(\vec{\phi} - \frac{v_i}{c}\vec{A}_i \right)\label{eq:wong}
    \end{eqnarray}
\end{subequations}
Here, $P_i$ and $v_i$ denote the particle momentum and velocity, respectively. 

\subsection{Maximally Non-Abelian Fields}
Equation~\eqref{eq:fdef} contains two qualitatively distinct contributions. The derivative terms represent Abelian-like components, while the term proportional to $g$ arises purely from the non-Abelian gauge structure. The fields that are entirely produced by the term proportional to $g$ shall be called \textit{``maximally non-Abelian''}.

The fields are maximally non-Abelian if, in a suitable gauge, the gauge potentials can be expressed as 
\begin{eqnarray}
    \vec{\phi} = -\frac{1}{c} \partial_0 \vec{\chi}, \qquad \vec{A}_i = \partial_i \vec{\chi}
\end{eqnarray}
The four curl of the above potentials is zero, and hence can not produce an Abelian-type field. In this gauge, the color electric and color magnetic fields are expressed as
\begin{subequations}
    \label{eq:maxnb-fdef}
    \begin{eqnarray}
        \vec{E} _i &=& \frac{g}{c} \left(\partial_0 \vec{\chi}\right) \times  \left(\partial_i \vec{\chi}\right) \label{eq:maxnb-e} \\
        \vec{B}_i &=& -g \left(\partial_j \vec{\chi}\right) \times \left(\partial_k \vec{\chi}\right) \label{eq:maxnb-b}
    \end{eqnarray}
\end{subequations}

Further, these gauge fields and the associated fields can be constants, which is the configuration we are interested in for this study.

\subsection{Classification of Field Configurations}
Now that the maximally non-Abelian fields are defined, we shall classify them into different configurations. 

\subsubsection{\label{subsec:1c-bf}One Component Magnetic Field}
The simplest field configuration is a color magnetic field with a single spatial and color component. This field can be generated with the following choice of gauge fields
\begin{eqnarray}
    \label{eq:1comp-a}
    \vec{A}_x =  A_x \hat{e}_2, \qquad \vec{A}_y =  A_y \hat{e}_1, \qquad \vec{\phi} = \vec{A}_z =0
\end{eqnarray}
The associated color magnetic field is
\begin{eqnarray}
    \label{eq:1comp-b}
    \vec{B}_i = gA_xA_y \hat{e}_3 \equiv B \hat{e}_3 \delta_{iz} 
\end{eqnarray}
The gauge potentials that generate this field are not unique. Consider this scale transformation
\begin{equation}
    \label{eq:scaling}
    A_x \rightarrow  \rho A_x, \qquad A_y \rightarrow \rho^{-1} A_y,
\end{equation}
under which the magnetic field is unchanged. However, each set of gauge potentials characterized by the parameter $\rho$ is gauge inequivalent to others, which means they are produced by different sources. This non-uniqueness in the gauge source is Wu-Yang ambiguity \cite{WuYang}, a characteristic feature of non-Abelian gauge theories.

The field configuration is widely and most easily realised in many systems, such as the spin-orbit coupling in spintronics systems, in ultracold atom laser systems, and in photonic systems with anisotropy. We describe one particular example arising in spintronics.\\

\noindent{\textit{Yang-Mills fields in Rashba-Dresselhaus Spin-orbit coupling:}} 
We consider a 2D system with a spin orbit coupling which has contributions from both the Rashba and the Dresselhaus types. The standard Hamiltonian has the form
\begin{eqnarray}
    \label{eq:r-h1}
    H = \frac{p^2}{2m} +  \left(g_1 p_y s_x - g_2 p_x s_y \right)
\end{eqnarray}
where the spin $\vec{s}$ of the electron acts as the color charge, $g_1$, $g_2$ are coupling strengths along the two coordinate directions and are system dependent and generally tunable\cite{Nitta1997}. From the above Hamiltonian, one can extract the non-Abelian potentials, which are of the form\cite{YMSpinTAN}
\begin{eqnarray}
    \label{eq:r-gf}
    \vec{A}_x = - \frac{mcg_2}{e} s_y \hat{e}_1, \qquad \vec{A}_y = \frac{mcg_1}{e} s_x \hat{e}_2
\end{eqnarray}
where, $e$ is the electric charge. The associated magnetic field is
\begin{eqnarray}
    \label{eq:r-gf}
    \vec{B}_z =  g~g_1g_2\frac{m^2c^2}{e^2} s_z \hat{e}_3
\end{eqnarray}
The medium could be engineered such that $g_1 \rightarrow \lambda g_1$ and $g_2 \rightarrow \lambda^{-1} g_2$, such that the color magnetic field is the same, exhibiting Wu-Yang ambiguity. 

\subsubsection{\label{subsec:3c-bf}Three Component Magnetic Field}
A more general case arises when all the three components of the vector potential survive. Since all the components are constant, we can perform a global bi-orthogonal transformation- a singular value decomposition (SVD) \cite{bhatia97}, to bring it to the canonical form. In this SVD basis, the gauge potential has the form
\begin{equation}
    \label{eq:3comp-a}
    \vec{A}_i = \left( A_x \hat{e}_1,\; A_y \hat{e}_2,\; A_z \hat{e}_3 \right)
\end{equation}
The resultant magnetic field has the same form as the potential and is given by
\begin{eqnarray}
    \label{eq:3d-bdef}
    \vec{B}_x = g A_y A_z \hat{e}_1, ~~ \vec{B}_y = g A_z A_x \hat{e}_2, ~~ \vec{B}_z = g A_x A_y \hat{e}_3
\end{eqnarray}
From the above equation, it can be seen that this particular field configuration does not exhibit Wu-Yang ambiguity, unlike the previous configuration. Also, from this equation, we can see that a planar magnetic field is not supported. This configuration has not been extensively studied experimentally; however, there have been proposals to realise this configuration in ultra cold atomic systems, which we shall briefly describe below as an example. 

\noindent{\textit{Three component field in cold atom system:}}
In ~\cite{anderson2012synthetic}, it was proposed that three-dimensional spin–orbit coupling can be engineered in $^{87}\mathrm{Rb}$ atoms. In this scheme, a set of laser fields couples four internal atomic levels in a specific configuration, giving rise to an emergent non-Abelian gauge potential of the form
\begin{eqnarray}
    \vec{A}_i = \cos{\theta_L} \frac{\kappa_{\perp}}{2}\left(\sigma_x\hat{e}_x + \sigma_y \hat{e}_y \right) + \sin\theta_L \kappa_{\parallel} \sigma_z \hat{e}_z
\end{eqnarray}
where, $\theta_L, \kappa_{\perp}$ and $\kappa_{\parallel}$ are determined by the laser parameters. By appropriately tuning these parameters, one can realise an isotropic gauge field configuration.

A similar form of three-dimensional spin–orbit coupling can also be implemented in ultracold atomic systems using inhomogeneous magnetic pulses \cite{anderson2013magnetically}, which likewise generate synthetic non-Abelian gauge fields. These systems, in principle, can be used to emulate the high-energy systems.

\subsubsection{\label{subsec:config1}Combined Color electric and Magnetic fields}
A combination of color electric and color magnetic fields arises when both scalar and vector potentials are present. This combination of fields generated by \eqref{eq:maxnb-fdef} is bi-orthogonal to each other. Which means, if we freeze the color degree of freedom and take the inner product in the real space, the product turns out to be zero. Similarly, an inner product in the color space with a fixed spatial degree also vanishes. \begin{eqnarray}
    \label{eq:edotb}
    \sum_{i} E^a_i B^a_i = 0, \qquad \sum_{a} E^a_i B^a_i = 0
\end{eqnarray}
In the above equation, superscript $a$ represents the color space components.\\

\noindent{\textit{Case \RNum{1}}:}\\
Consider the field configuration that results in a one component color magnetic field \eqref{eq:1comp-a}; to this configuration, we introduce a scalar potential such that it is parallel to one of the vector potential components. The field configuration is as follows
\begin{equation}
    \label{eq:comb-case1-a}
    \vec{A}_x = A_x \hat{e}_2, \qquad
    \vec{A}_y = A_y \hat{e}_1, \qquad
    \vec{\phi} = \phi \hat{e}_2 .
\end{equation}
The associated color fields are
\begin{equation}
    \label{eq:fc-case1}
    \vec{E}_y =  g \phi A_y \hat{e}_3 \qquad
    \vec{B}_z = g A_x A_y \hat{e}_3
\end{equation}
while all other components vanish. The resulting fields have one spatial and color component each.\\

\noindent{\textit{Case \RNum{2}}:}\\
Next, to the one-component field configuration \eqref{eq:1comp-a}, a scalar potential is introduced such that it is perpendicular to both the vector potential components in the color space. The field configuration is as follows
\begin{eqnarray}
    \label{eq:comb-case2-a}
    \vec{A}_x = A_x \hat{e}_2, \qquad
    \vec{A}_y = A_y \hat{e}_1, \qquad
    \vec{\phi} = \phi \hat{e}_3
\end{eqnarray}
This generates the following color magnetic and electric fields:
\begin{eqnarray}
    \label{eq:fc-case2}
    \vec{B}_z &=& g A_x A_y \hat{e}_3, \qquad
    \vec{E}_x = g A_x \phi \hat{e}_1, \\ \nonumber
    \vec{E}_y &=& -g A_y \phi \hat{e}_2.
\end{eqnarray}
while all remaining field components vanish.\\

Case \RNum{1} and \RNum{2} are of special interest, and these combinations of fields can possibly be realised in the near future as proposed in \cite{chen2019non}, for example. In this proposal, the permittivity and permeability of the material are altered in the presence of an electromagnetic wave to achieve the desired configurations. Finally, the dynamics of a test particle in the most general combination of color electric and magnetic field is discussed in the Appendix~\ref{app:config3} for completeness.\\ 

\section{\label{sec:Magdyn-1D}Particle Motion in One-Component Color Magnetic Field}
Before we discuss the dynamics of the test particle, we shall introduce the units employed in this study.

\subsection{\label{sec:units}Units Employed}
In this work, we employ a set of units furnished by the system. For ready comparison with the ED, we introduce the scaled charge $Q \equiv g q$, which has the dimensions of the electric charge. The other dimensionless variables are
\begin{eqnarray}
    \tau &=& \left(\frac{QB}{mc}\right)t, \qquad p_i = \frac{P_i}{mc}, \nonumber \\ 
    u_i &=& \frac{v_i}{c},  \qquad \varepsilon = \frac{\mathcal{E}}{mc^2}, \qquad\vec{\zeta} = \frac{\vec{Q}}{Q} \nonumber  \\ 
    \vec{a}_i,\vec{\varphi} &=& \left(\frac{Q}{\kappa mc^{2}}\right)\vec{A}_i,\vec{\phi}, \qquad \vec{b}_i = \frac{\vec{B}_i}{B}. 
\end{eqnarray}
Here, $B$ is the magnitude of the color magnetic field, and
\begin{equation}
    \kappa = \frac{gq^{2} B}{\left(mc^{2}\right)^{2}}.
\end{equation}
The constant $\kappa$ compares the energy scale associated with non-Abelian magnetic field interaction to the rest mass energy of the particle. The parameter $\kappa$ therefore plays the role of an effective coupling constant governing the strength of the gauge interaction.

In terms of these dimensionless variables, the equations of motion (\ref{eq:dyn}) in the presence of a color magnetic field take the form
\begin{subequations}
    \label{eq:neom}
    \begin{eqnarray}
        \frac{dp_i}{d\tau} &=& \vec{\zeta} \cdot \epsilon_{ijk} u_j\vec{b}_k  \label{eq:nlz} \\
        \frac{d\vec{\zeta}}{d\tau} &=& -\vec{\zeta} \times \vec{a}_i u_i \label{eq:nwong}
    \end{eqnarray}
\end{subequations}
The canonical momentum of the particle is conserved as a consequence of the constancy of gauge potentials. In addition, the magnitude of the color charge is conserved, as can be seen from the equation \eqref{eq:nwong}. This conservation is a direct consequence of the symmetry under consideration, where the magnitude of the charge turns out to be the Casimir of the group. Before we discuss the particle dynamics, it should be noted that most physical realisations lie in the non-relativistic regime; we retain the relativistic form of the equations for the sake of completeness; the non-relativistic limit follows straightforwardly. With the equations of motion expressed entirely in dimensionless form, we now proceed to analyse the resulting particle dynamics.

\subsection{Test Particle Dynamics}
As discussed in Sec.~\ref{sec:classical-YM}, the simplest realisation of a maximally non-Abelian magnetic field is the one-component magnetic field, which points along the $z$ direction in real space and along $\hat{e}_3$ in the color space. For this field configuration, the dimensionless equations of motion (\ref{eq:neom}) reduce to
\begin{eqnarray}
    \frac{d u_x}{d \tau} &=& \frac{u_y \zeta_3}{\gamma}; ~ \frac{d u_y}{d \tau} = -\frac{u_x \zeta_3}{\gamma}; \nonumber \\
    \frac{d \zeta_1}{d \tau} &=& a_x u_x \zeta_3; ~\frac{d \zeta_2}{d \tau} = -a_y u_y \zeta_3; \nonumber\\
    \frac{d \zeta_3}{d \tau} &=&  a_y u_y \zeta_2 - a_x u_x \zeta_1
\end{eqnarray}
since the motion along $z-$ direction is not affected by the field, we set $u_z = 0$, which remains preserved throughout the evolution. For the configuration under consideration, the components of the canonical momentum are
\begin{subequations}
    \label{eq:pi}
    \begin{eqnarray}
        \Pi_x = \gamma u_x + a_x \zeta_2\\
        \Pi_y = \gamma u_y + a_y \zeta_1
    \end{eqnarray}
\end{subequations}
These relations imply that the evolutions of the velocity $u_i (\tau)$  and the color charge $\vec{\zeta}(\tau)$ follow each other. Additionally, the kinetic energy of the particle is conserved. We parameterise the components of velocity as
\begin{eqnarray}
    \label{eq:para}
    u_x(\tau) &=& u \cos\theta(\tau), \qquad ~ u_y = u \sin\theta(\tau)
\end{eqnarray}
where $\theta(\tau)$ denotes the instantaneous direction of motion in the plane. In terms of this parametrisation, the equation governing the motion of $\theta(\tau)$ takes the form 
\begin{widetext}
    \begin{eqnarray}
        \label{eq:thetadot}
        \frac{d \theta(\tau)}{d \tau} = \pm \frac{1}{\gamma} \sqrt{1 - \left \{ \gamma a_x \left (\sin{\theta(0)} - \sin{\theta}\right) +\zeta_1(0) \right \}^2 - \left \{\gamma a_y \left(\cos{\theta(0)} - \cos{\theta}\right) + \zeta_2(0) \right\}^2} 
    \end{eqnarray}
\end{widetext}
where $\theta(0)$, $\zeta_1(0)$, and $\zeta_2(0)$ denote the initial values. 

Equation~\eqref{eq:thetadot} admits closed-form solutions for the special case when $a_x = a_y =a~~ \text{and}~~\zeta_3(0) = 1$. Before presenting closed-form solutions, it is instructive to examine the qualitative implications of Eq.~\eqref{eq:thetadot}. In ED, the angular variable evolves linearly, $\theta(\tau) = \omega_c \tau$, resulting in closed cyclotron orbits ($\omega_c$ is the gyration frequency). In contrast, the non-Abelian coupling leads to a nonlinear evolution of $\theta(\tau)$. This nonlinearity modifies the radial displacement of the particle and can produce unbounded trajectories. The resulting drift resembles that of a charged particle in a time-dependent magnetic field, but here it arises purely from the dynamical evolution of the color charge. The nuances of the dynamics are sensitive to the kinetic energy, the magnitude of the gauge potentials, and the initial conditions.\\

\noindent{\textit{Symmetric Case} $(a_x = a_y = a; \zeta_3(0) = 1)$:} The nonlinearity of $\theta(\tau)$ is made explicit in the symmetric case, on plugging in the conditions into the Eq.~\eqref{eq:thetadot}, it reduces to: 
\begin{eqnarray}
    \frac{d \theta}{d \tau} = \pm\frac{1}{\gamma} \sqrt{1 - 4\gamma^2 u^2 a^2 \sin^2 \left ( \frac{\theta - \theta(0)}{2} \right)}, ~ k = 2\gamma u a,
\end{eqnarray}
the solution to the above equation is given by
\begin{eqnarray}
    \label{eq:thetasol}
    \theta(\tau) = \theta(0) +   {\mathbf{{am}}}\left(-\frac{b \tau}{2 \gamma}, k^2 \right)
\end{eqnarray}
where $\mathbf{am}$ denotes the Jacobi amplitude function \cite{whittaker2020course}. The above solution is the same as the solution of pendulum motion without the small-angle approximation (the value of $k$ in the case of the pendulum depends on the initial angle). The corresponding color charge and velocity components evolve as
\begin{widetext}
    \begin{subequations}
        \label{eq:sol1D}
        \begin{eqnarray}
            \zeta_1(\tau) &=& \frac{k}{2} \left[ \sin{(\theta(0))} - \sin\left ( \theta(0) +  \mathbf{am}\left(-\frac{\tau }{2\gamma}, k^2 \right) \right)\right]\\
            \zeta_2(\tau) &=& \frac{k}{2} \left[ \cos{(\theta(0))} - \cos\left(\theta(0) +  \mathbf{am}\left(-\frac{\tau }{2\gamma}, k^2 \right) \right)\right]\\
            \zeta_3(\tau) &=& \mp \gamma \mathbf{dn}\left ( -\frac{ \tau }{2\gamma}, k^2 \right)\\
            u_x(\tau) &=& \sqrt{2\varepsilon} \cos{\left(\theta(0) + {\mathbf{{am}}}\left(-\frac{\tau}{2 \gamma}, k^2 \right) \right)}\\
            u_y(\tau) &=& \sqrt{2\varepsilon} \sin{\left(\theta(0) + {\mathbf{{am}}}\left(-\frac{\tau}{2 \gamma}, k^2 \right) \right)}
        \end{eqnarray}
    \end{subequations}
\end{widetext}
where $\mathbf{dn}$ denotes the Jacobi delta function \cite{whittaker2020course}. This set of solutions presented in \eqref{eq:sol1D} also helps us to benchmark the numerical calculations. For $k \neq 0$, the evolution of $\theta(\tau)$ is non-linear, leading to periodic motion with a nonzero average displacement per cycle, $x_{mean},y_{mean} \neq \text{constant}$, and hence leads to a linear drift of the trajectory. In the limit $k \rightarrow 0$, the drift becomes vanishingly small, and the dynamics approach the Abelian limit. 

\begin{figure*}[t]
    \centering
    \begin{subfigure}{0.30\textwidth}
        \centering
        \includegraphics[width=\linewidth]{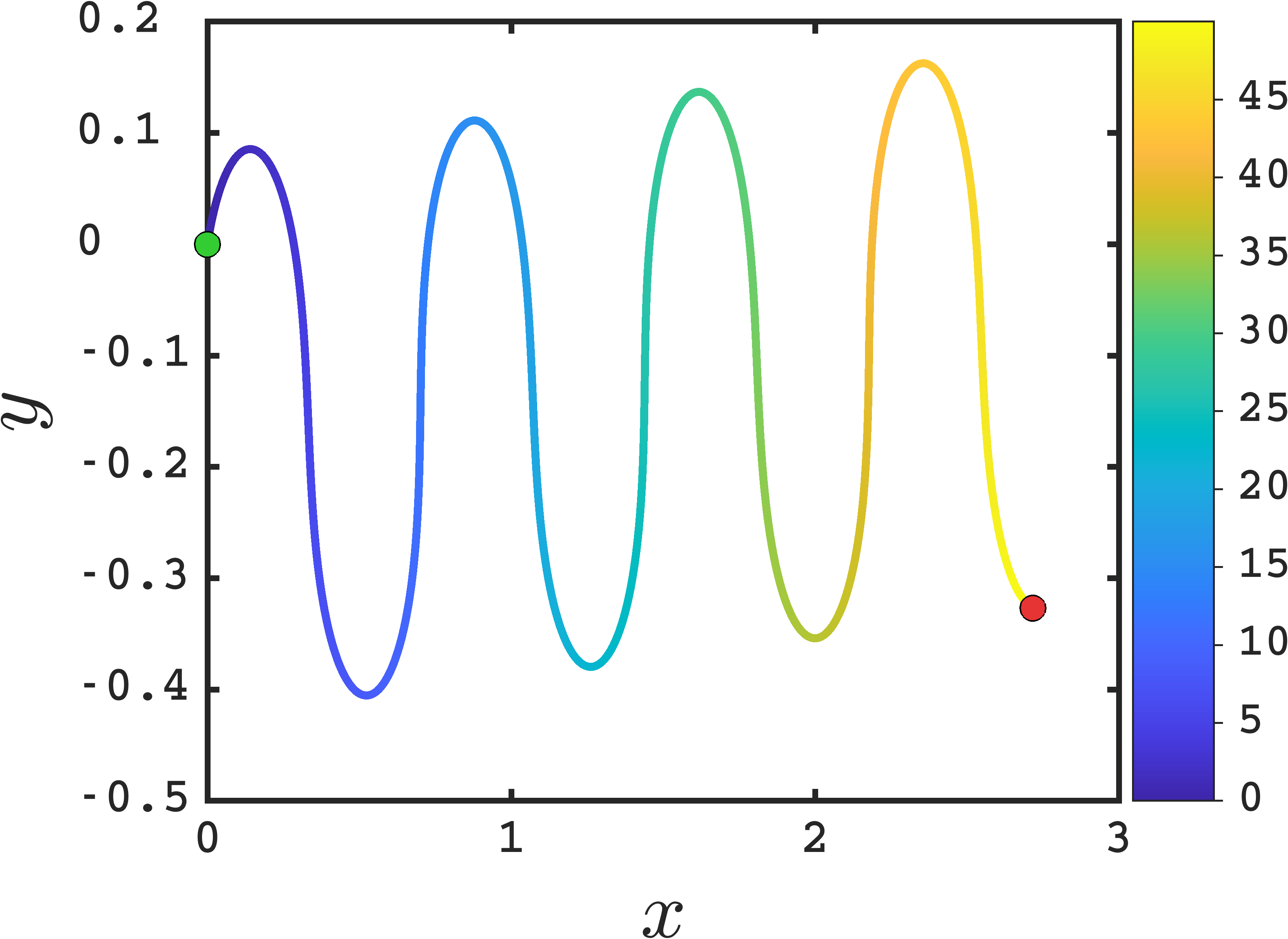}
        \caption{Trajectory for $\rho^{-1}=16$, $u(0)=0.1$, $\zeta_1(0)=0.61$, $\zeta_2(0)=0.61$, $\zeta_3(0)=0.5$.}
        \label{fig:1a-cla}
    \end{subfigure}
    \hfill
    \begin{subfigure}{0.30\textwidth}
        \centering
        \includegraphics[width=\linewidth]{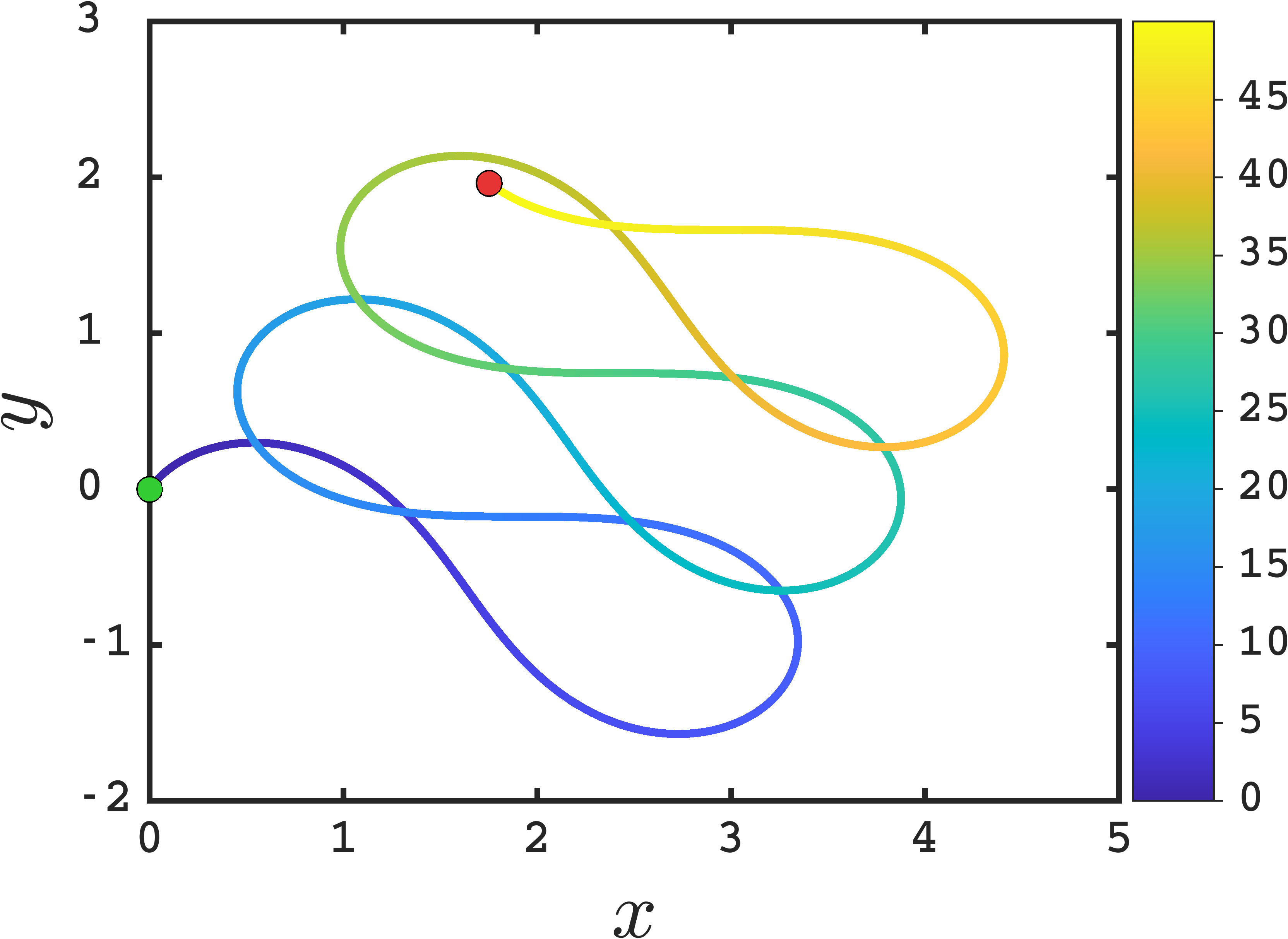}
        \caption{Trajectory for $\rho=1$, $u(0)=0.5$, $\zeta_3(0)=1$.}
        \label{fig:1c-cla}
    \end{subfigure}
    \hfill
    \begin{subfigure}{0.30\textwidth}
        \centering
        \includegraphics[width=\linewidth]{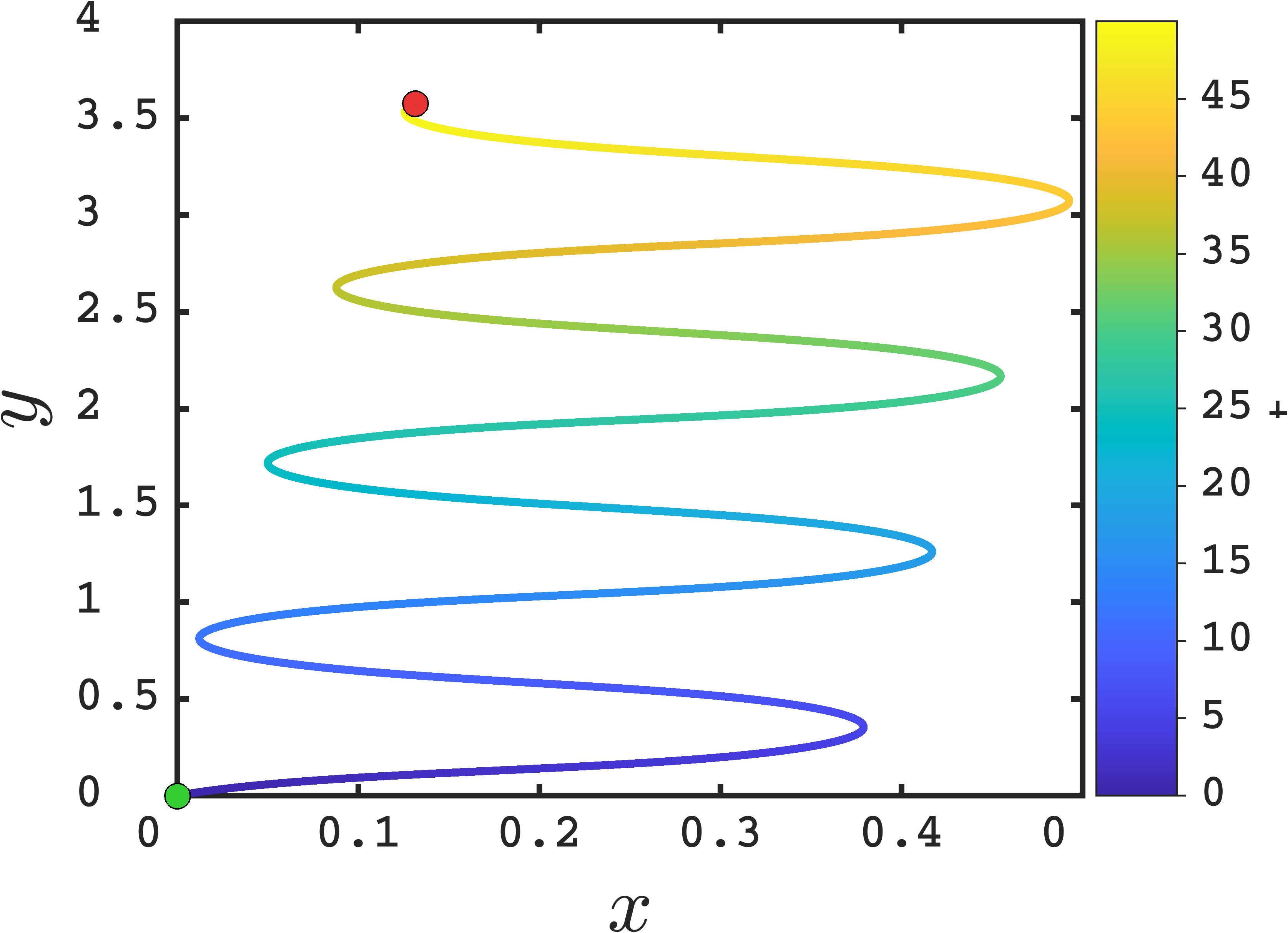}
        \caption{Trajectory for $\rho=16$, $u(0)=0.1$, $\zeta_1(0)=0.61$, $\zeta_2(0)=0.61$, $\zeta_3(0)=0.5$.}
        \label{fig:1b-cla}
    \end{subfigure}

    \medskip
    \begin{subfigure}{0.30\textwidth}
        \centering
        \includegraphics[width=\linewidth]{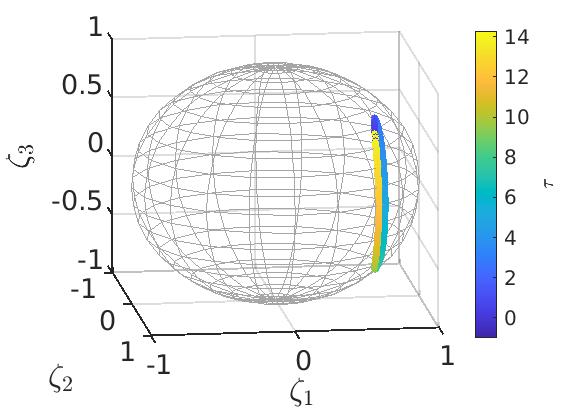}
        \caption{Color charge evolution corresponding to Fig. \ref{fig:1a-cla}.}
        \label{fig:2a-cd}
    \end{subfigure}
    \hfill
    \begin{subfigure}{0.30\textwidth}
        \centering
        \includegraphics[width=\linewidth]{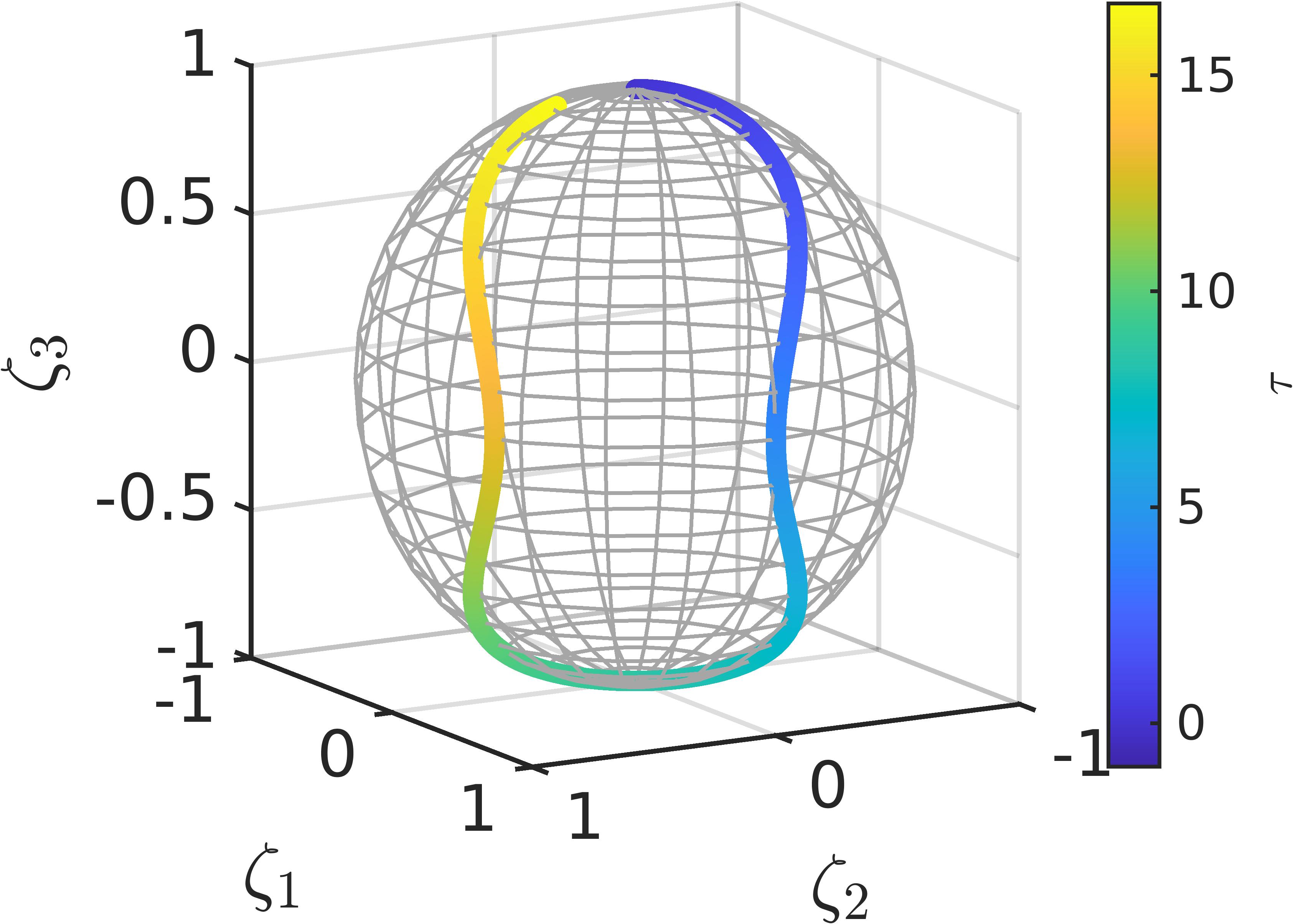}
        \caption{Color charge evolution corresponding to Fig. \ref{fig:1c-cla}.}
        \label{fig:2c-cd}
    \end{subfigure}
    \hfill
    \begin{subfigure}{0.30\textwidth}
        \centering
        \includegraphics[width=\linewidth]{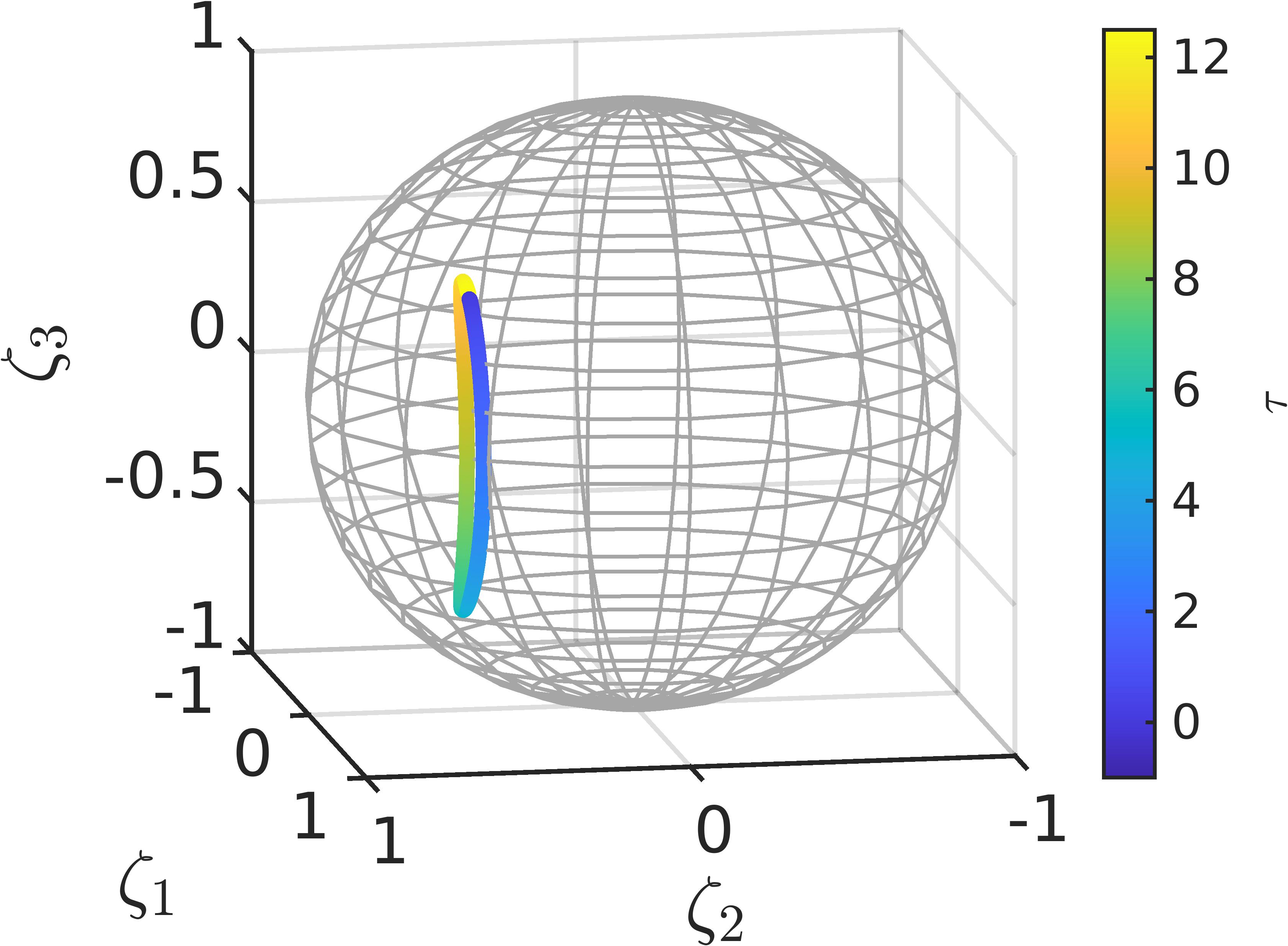}
        \caption{Color charge evolution corresponding to Fig. \ref{fig:1b-cla}.}
        \label{fig:2b-cd}
    \end{subfigure}
    \hfill
    \caption{Particle trajectories for different gauge potential configurations at fixed magnetic field strength. The color map represents time evolution. Panels below each trajectory show the corresponding evolution of the color charge on the internal phase space.}
    \label{fig:1-cla}
\end{figure*}

We now present the results of the numerical calculation for all choices of gauge potentials with the strength of the magnetic field fixed. For all choices of gauge potentials and initial conditions, we see a drift in the trajectory of the particle. Representative particle trajectories for different values of $\rho$ (both asymmetric and symmetric cases) are shown in Fig.~\ref{fig:1-cla}, all of which exhibit unbounded motion. The exotic trajectory displayed in Fig.~\ref{fig:1c-cla} arises from a dynamical reversal of the color charge direction during the evolution. The corresponding phase-space evolution of the color charge is shown in the lower panel of the same figure.

\begin{figure*}[t]
    \centering
    \includegraphics[width=0.45\linewidth]{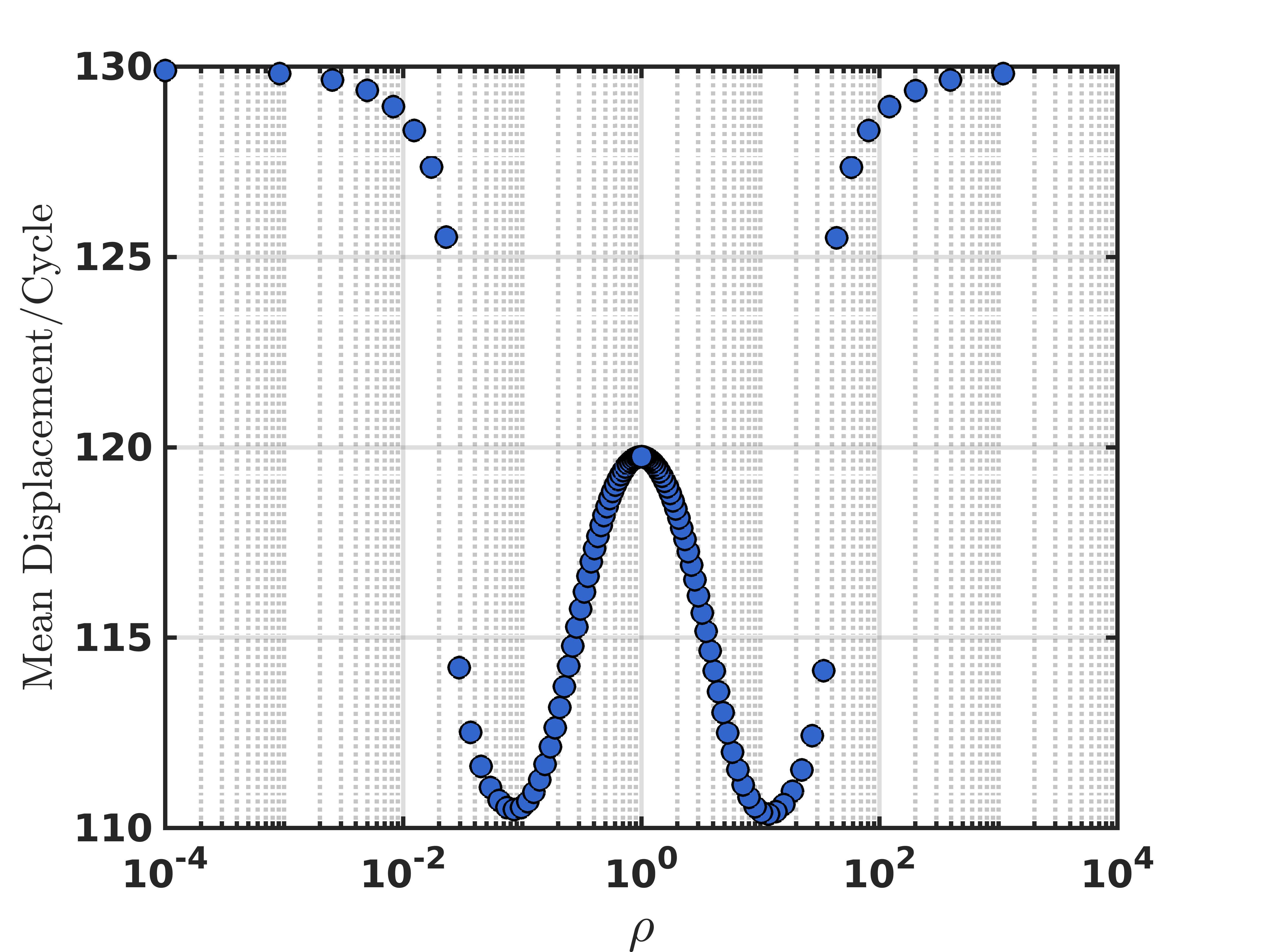}
    \caption{Illustrates the mean displacement per cycle as a function of $\rho$. The $x-$ axis of the plot is in $log$ scale. Here, the initial conditions are, $p(0) = 1.0$ with $p_x(0) = p_y(0)$ and $\zeta_1(0) = \zeta_2(0) =  \zeta_3(0) = 0.57$}
    \label{fig: drift_one_comp}
\end{figure*}

Further, in Fig. \ref{fig: drift_one_comp}, we illustrate how the mean displacement of the particle over a cycle varies as a function of $\rho$. As seen from the figure, this mean displacement saturates when one of the gauge potentials is much greater than the other. The symmetry in the plot under $\rho \rightarrow \rho^{-1}$ arises from the specific initial conditions, where magnitudes of both the velocity components are equal to each other and the magnitudes of the charge components are equal to each other. This symmetry is broken upon changing the initial conditions; however, the saturation persists due to the conservation of canonical momentum. As one can see from these plots, the total drift in the particle's trajectory is never zero.

The dynamics uncovered here has direct relevance to physical systems. From an experimental perspective, the currents generated in these systems need not follow closed cyclotron orbits even if the synthetic magnetic fields are uniform. The drift seen here can potentially be reflected in the behaviour of transverse spin currents or spin-Hall responses.

Having analysed particle dynamics in a single-component constant color magnetic field, we now turn to the more general case of a three-component constant color magnetic field and investigate the particle motion.

\section{\label{sec:Magdyn-3D}Particle Motion in Three-Component Color Magnetic Fields}
The field configuration has been described in the subsection \ref{subsec:3c-bf}, in equations \eqref{eq:3comp-a}-\eqref{eq:3d-bdef}. For this field configuration, the set of equations in \eqref{eq:neom} yields a system of six coupled first order differential equations, governing the evolution of the particle’s velocity and color charge. 

The components of the canonical momentum for the present configuration are
\begin{eqnarray}
        \Pi_x &=& \gamma u_x + a_x \zeta_1,\qquad \Pi_y = \gamma u_y + a_y \zeta_2 \\ \nonumber
        \Pi_z &=& \gamma u_z + a_z \zeta_3
\end{eqnarray}

Since the evolution of the color degrees of freedom $\vec{\zeta}(\tau)$ follows the dynamics of the momenta, these relations allow $\vec{\zeta}(\tau)$ to be eliminated in favour of the velocity components $u_i(\tau)$. This reduction yields a closed system of three coupled equations for the components of velocity:
\begin{subequations}
    \label{eq:rdeom}
    \begin{eqnarray}
        \frac{d u_x}{d \tau} &=& \frac{1}{\gamma}\left(u_z \mathcal{B}_y  - u_y \mathcal{B}_z  \right) + u_y u_z \left(\frac{b_z}{a_z} - \frac{b_y}{a_y} \right)\\
        \frac{d u_y}{d \tau} &=& \frac{1}{\gamma}\left(u_x \mathcal{B}_z  - u_z \mathcal{B}_x  \right) + u_z u_x \left(\frac{b_x}{a_x} - \frac{b_z}{a_z} \right)\\
        \frac{d u_z}{d \tau} &=& \frac{1}{\gamma} \left(u_y \mathcal{B}_x  - u_x \mathcal{B}_y  \right) + u_x u_y \left(\frac{b_y}{a_y} - \frac{b_x}{a_x} \right)
    \end{eqnarray}
\end{subequations}
where
\begin{eqnarray}
    \mathcal{B}_x &=& \frac{b_x \Pi_x}{a_x},\qquad \mathcal{B}_y = \frac{b_y \Pi_y}{a_y}, \qquad \mathcal{B}_z = \frac{b_z \Pi_z}{a_z}
\end{eqnarray}
The system of equations \eqref{eq:rdeom} is numerically integrated. In general, the solutions are nonlinear and lead to drifting particle trajectories. 

A particularly instructive special case arises when the magnetic field components are equal,
\begin{eqnarray}
    a_x = a_y = a_z = a ~~ \implies ~b_x = b_y = b_z = b 
\end{eqnarray}
Under this condition, the equations of motion simplify to
\begin{eqnarray}
    \frac{d u_i}{d \tau} &=& - \frac{1}{\gamma}\epsilon_{ijk}u_j \mathcal{B}_k 
\end{eqnarray}
which is formally identical to the Lorentz equation for a particle moving in a uniform effective magnetic field ${\mathcal{B}}_i$. Importantly, this effective field is entirely determined by the conserved quantities and the gauge potentials. It is the only situation under which the particle undergoes cyclotron motion, and the trajectory is bounded. However, note that the radius of the trajectory and the frequency of the gyration are dependent on the canonical momentum and not just on the magnetic field and the gauge potentials. The frequency of gyration is given by $|\mathcal{B}_i|/\gamma$. By suitably choosing the initial conditions, it is possible to achieve $\mathcal{B}_i = 0$, in which case the particle experiences no net force and propagates freely, despite the presence of a nonzero non-Abelian magnetic field.

\begin{figure*}[t]
    \centering
    \begin{subfigure}{0.45\textwidth}
        \centering
        \includegraphics[width=\linewidth]{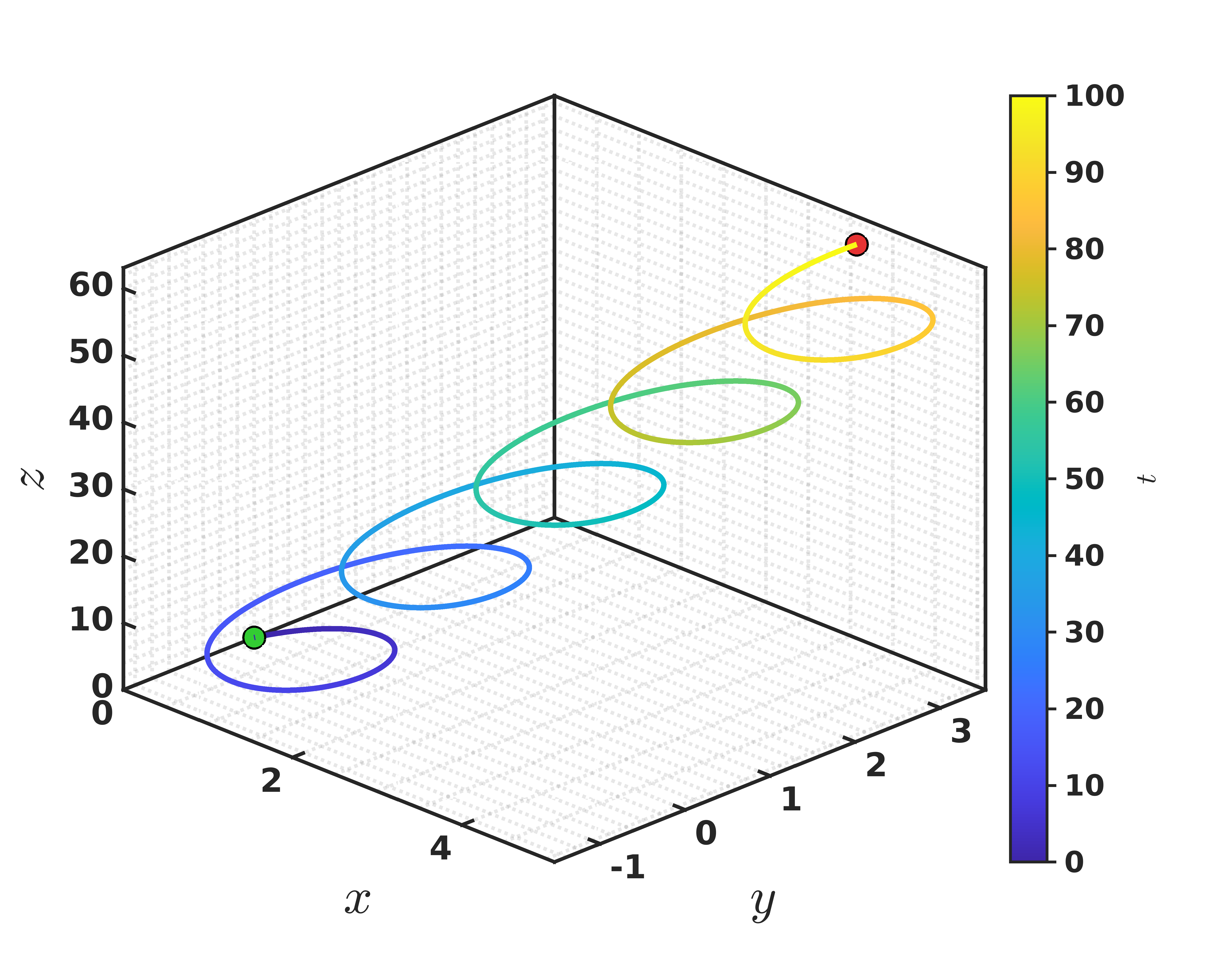}
        \caption{Trajectory in a general three-component non-Abelian magnetic field
        ($b_x=0.19$, $b_y=0.33$, $b_z=0.92$).}
        \label{fig:3d-2a}
    \end{subfigure}
    \hfill
    \begin{subfigure}{0.45\textwidth}
        \centering
        \includegraphics[width=\linewidth]{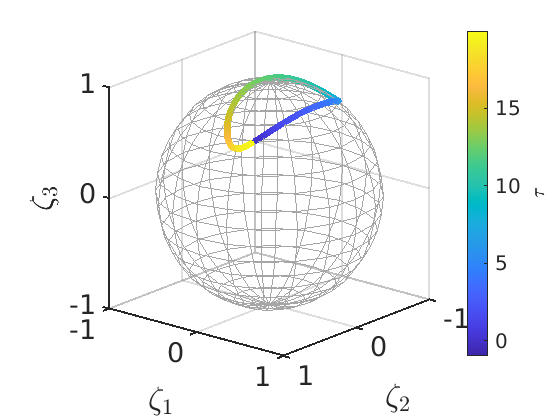}
        \caption{Corresponding evolution of the color charge in internal phase space.}
        \label{fig:3d-2a-coldy}
    \end{subfigure}
    \caption{Particle trajectory (left panel) and corresponding color charge evolution in internal phase space (right panel) for three-component non-Abelian magnetic fields. The color map represents time evolution.}
    \label{fig:2}
\end{figure*}

\begin{figure*}[t]
    \centering
    \begin{subfigure}{0.3\textwidth}
        \centering
        \includegraphics[width=\linewidth]{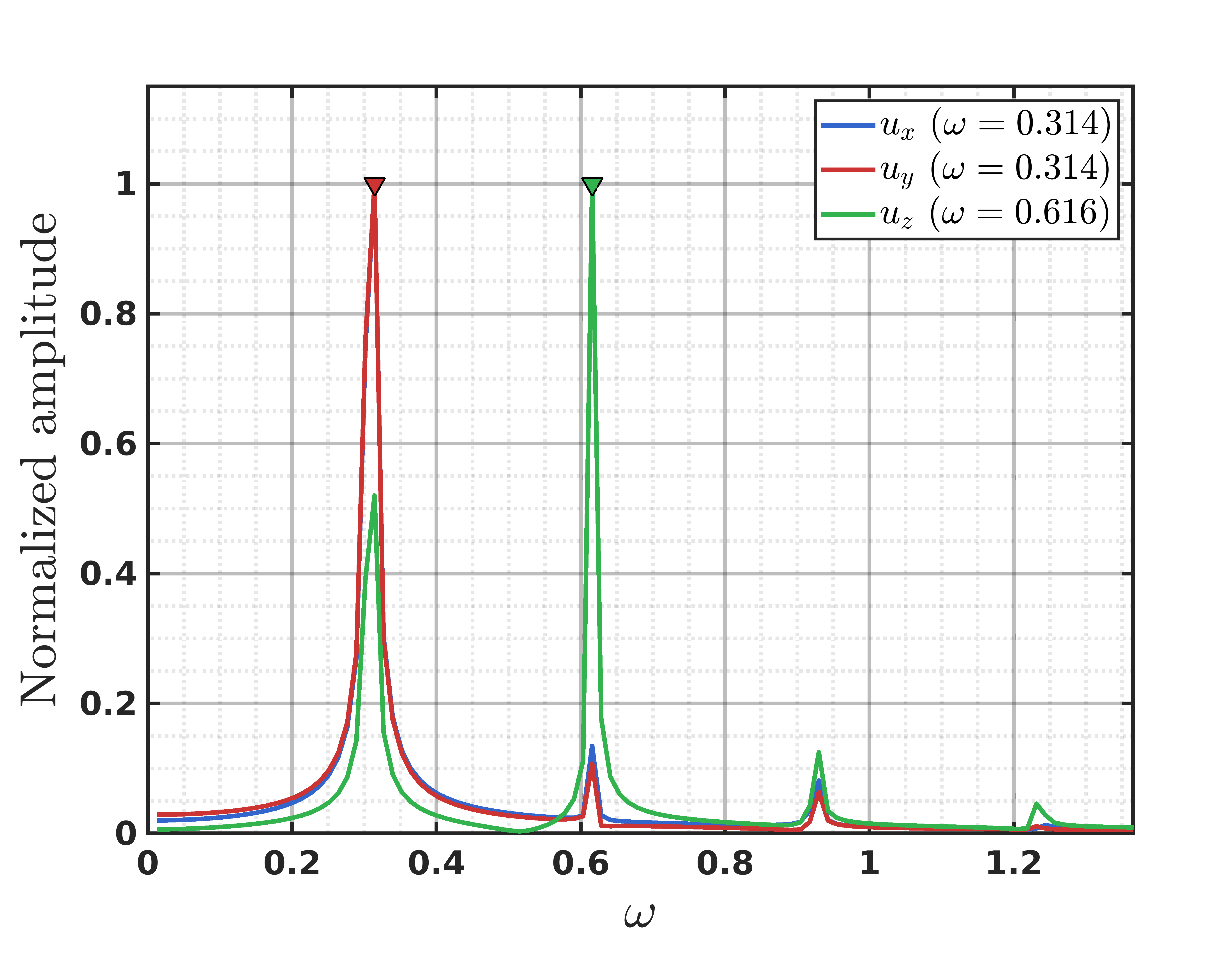}
        \caption{Frequency spectrum of different components of velocity, when $b_x=0.19$, $b_y=0.33$,$b_z=0.92$ and $\zeta_1(0) = \zeta_2(0) = 0.41, \zeta_3(0) = 0.80$. The evolution has multiple time periods.}
        \label{fig:fft-1a}
    \end{subfigure}
    \hfill
    \begin{subfigure}{0.3\textwidth}
        \centering
        \includegraphics[width=\linewidth]{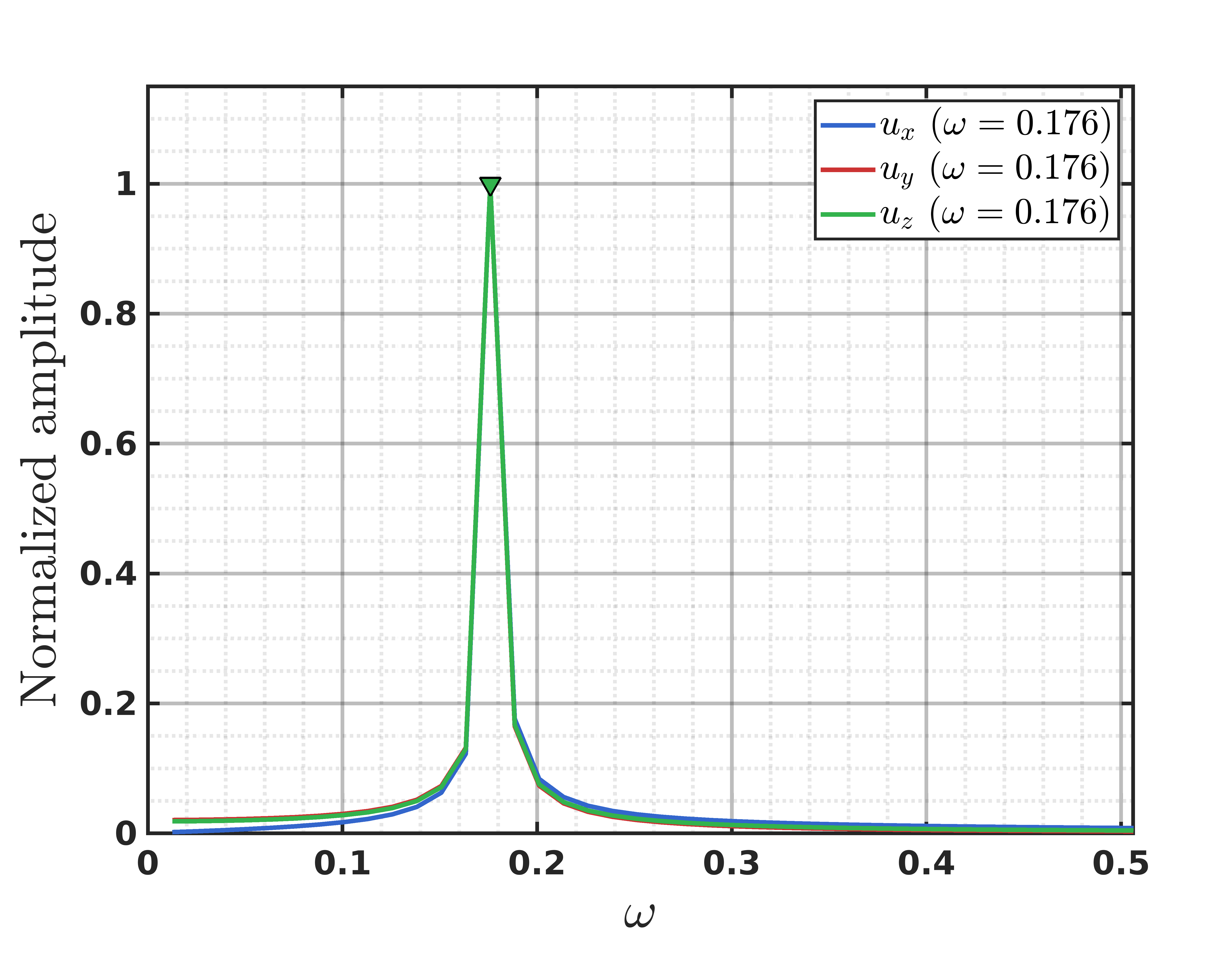}
        \caption{Frequency spectrum of different components of velocity, when $b_x= b_y= b_z=0.57$and $\zeta_1(0) = \zeta_2(0) = 0.41, \zeta_3(0) = 0.80$. The evolution has a single time period, given by the effective field $\mathcal{B}_i$, which is $\omega_{(ana)} = 0.17$.}
        \label{fig:fft-1b}
    \end{subfigure}
    \hfill
    \begin{subfigure}{0.3\textwidth}
        \centering
        \includegraphics[width=\linewidth]{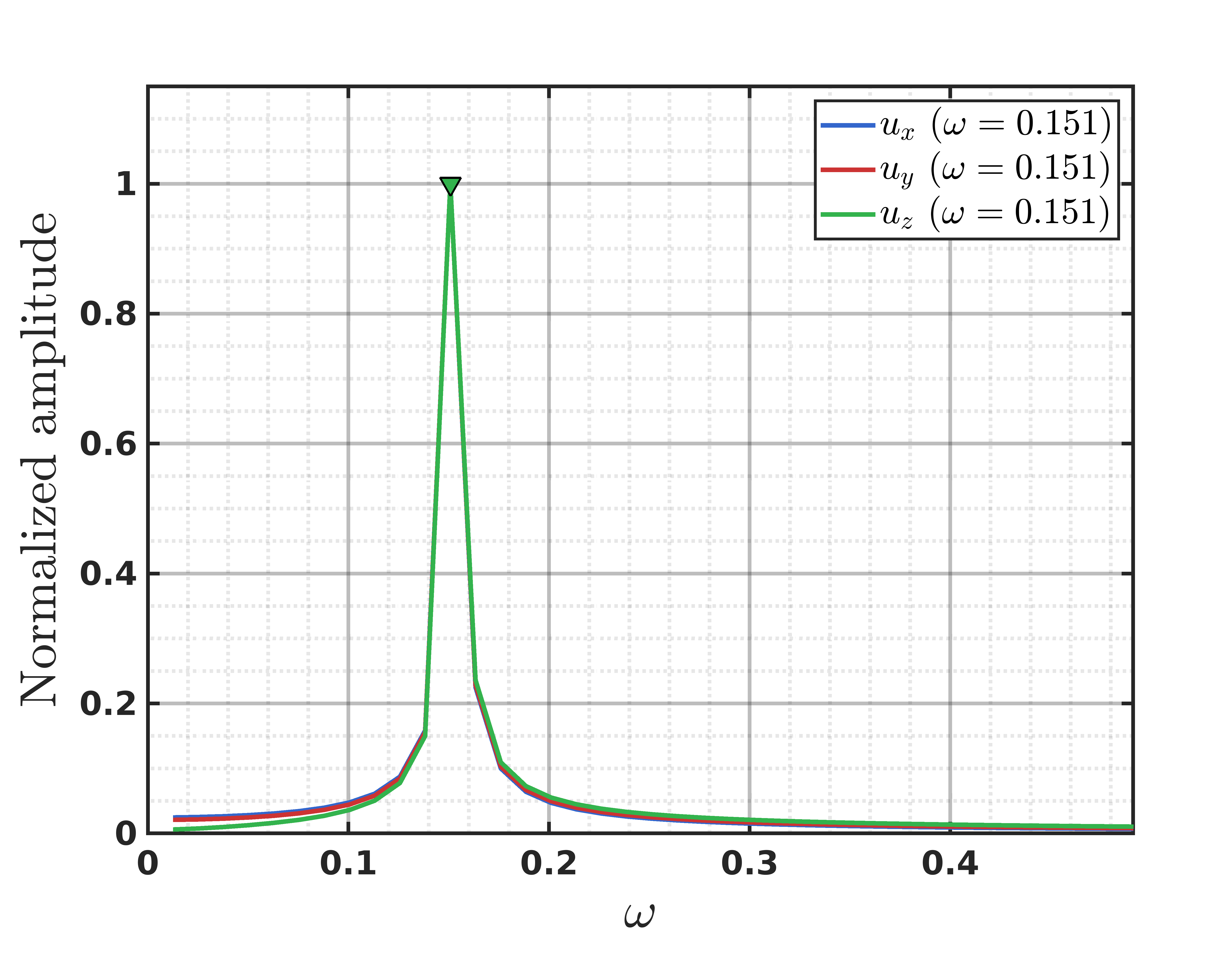}
        \caption{Frequency spectrum of different components of velocity, when $b_x= b_y= b_z=0.57$ and $\zeta_1(0) = \zeta_2(0) = 0.41, \zeta_3(0) = 0.80$. The evolution has a single time period, given by the effective field $\mathcal{B}_i$, which is $\omega_{(ana)} = 0.15$.}
        \label{fig:fft-1c}
    \end{subfigure}
    \caption{Shows the frequency spectrum of particle's velocity components for different choices of gauge potentials and initial conditions in the three component magnetic field configuration.}
    \label{fig:fft-1}
\end{figure*}

A representative trajectory illustrating the drift in a most general color magnetic field is shown in Fig.~\ref{fig:2}. The left panel displays motion in a generic three-component non-Abelian magnetic field, characterised by nonlinear dynamics and a drifting trajectory, along with the phase space evolution of the color charge on the right. In Fig. \ref{fig:fft-1}, we show the multiple time scales involved in the evolution of the particle by plotting the frequency spectrum of the components of its velocity. It can be seen that, for a general field configuration, the evolution has multiple time scales, but when $a_x = a_y = a_z$, the evolution has a single time scale. In this symmetric case, the frequency is given by the magnitude of the effective field $\vert \mathcal{B}_i \vert$. In the same figure, the dependence of gyration frequency on the initial conditions for the symmetric case is shown. Since we do not have scaling as in the case of the one-component configuration, we do not show how the drift varies for different choices of the gauge fields.

\section{\label{sec:Combdyn}Particle Motion in Combined Color Electric and Magnetic Fields}
In the preceding sections, we analysed the dynamics of a test particle in constant color magnetic fields. Many physical systems have scalar potential along with the vector potential. Thus, the combination of both color electric and color magnetic fields becomes a relevant case. A systematic study of particle motion in such combined fields is therefore essential for developing a complete understanding of test particle dynamics in the presence of synthetic non-Abelian gauge fields.

In ED, the combination of electric and magnetic fields plays a central role in plasma and transport phenomena. In particular, mutually perpendicular electric and magnetic fields give rise to the well-known $E \times B$ drift, which governs bulk particle transport in magnetised plasmas. In a non-Abelian setting, the situation is fundamentally richer due to the coupling between the color charge and momentum. Studying particle dynamics in combined color electric and magnetic backgrounds, therefore, provides a natural framework for contrasting YM dynamics with its Abelian counterpart.

In this section, we continue to use the units introduced previously. Here, we choose to express the magnitude of the color electric field as a ratio of the magnitude of the colour magnetic field, that is $\alpha_{i} = E_{i}/B$.  

\subsection{\label{sec:config1}Particle Motion in Field Configuration- Case \RNum{1}}
The field configuration is described in the subsection \ref{subsec:config1} in equation \eqref{eq:fc-case1} .Here, the color electric and magnetic fields are perpendicular to each other in real space, but parallel in the internal space. In the presence of a scalar potential, although the kinetic energy of the particle is not conserved, it is bounded due to the conservation of the total energy. The equations of motion specific to this configuration follow directly from the normalised version of Eqs.~\eqref{eq:dyn}. Since the $z$ component of the momentum is unaffected by the fields, we set $p_z = 0$ without loss of generality. 

The conserved quantities for this configuration are
\begin{eqnarray}
    \Pi_x &=& p_x + \zeta_2 a_x, \qquad \Pi_y = p_y + \zeta_1 a_y, \nonumber \\ 
    \varepsilon &=& \gamma + \zeta_2 \varphi
\end{eqnarray}

These constants of motion impose constraints on the particle dynamics. The conserved canonical momentum and the total energy restrict the allowed range of the particle's velocity. Using $\vert \zeta_i(\tau) \vert \leq 1$, the constraints on the kinetic energy is given  by the set of equations

\begin{eqnarray}
    \label{eq:pi-eb1}
   (\Pi_x - a_x) &\leq& p_x \leq (\Pi_x + a_x), \nonumber \\  
   (\Pi_y - a_y) &\leq& p_y \leq (\Pi_y + a_y) \nonumber \\ 
   (\varepsilon - \varphi) &\leq& \sqrt{1+p^2}\leq (\varepsilon + \varphi)
\end{eqnarray}

Within these bounds, the particle motion is nonlinear.

Alternatively, note that it is possible to perform a Lorentz transformation that eliminates either the color electric field or the color magnetic field, depending on whether $\alpha_{y} > 1$ or $\alpha_{y} < 1$, respectively. This is possible because the transformation preserves the internal color direction of the fields. When $\alpha_y > 1$, the resulting dynamics in the transformed frame is given by the results discussed in our earlier work \cite{KN_2025}. When $\alpha_y < 1$, the dynamics of the particle in the transformed frame is as described in section \ref{sec:Magdyn-1D} of this paper.  As a result, this configuration provides a convenient setting for analysing drift contributions associated with an $E \times B$–type mechanism from those arising purely due to non-Abelian electric or magnetic effects. The resulting dynamics offers a clear point of comparison between Abelian and non-Abelian drift phenomena in the field that obey $\vec{E}_i \cdot \vec{B}_i = 0$.

\begin{figure*}[t]
    \centering
    \begin{subfigure}{0.45\textwidth}
        \centering
        \includegraphics[width=\linewidth]{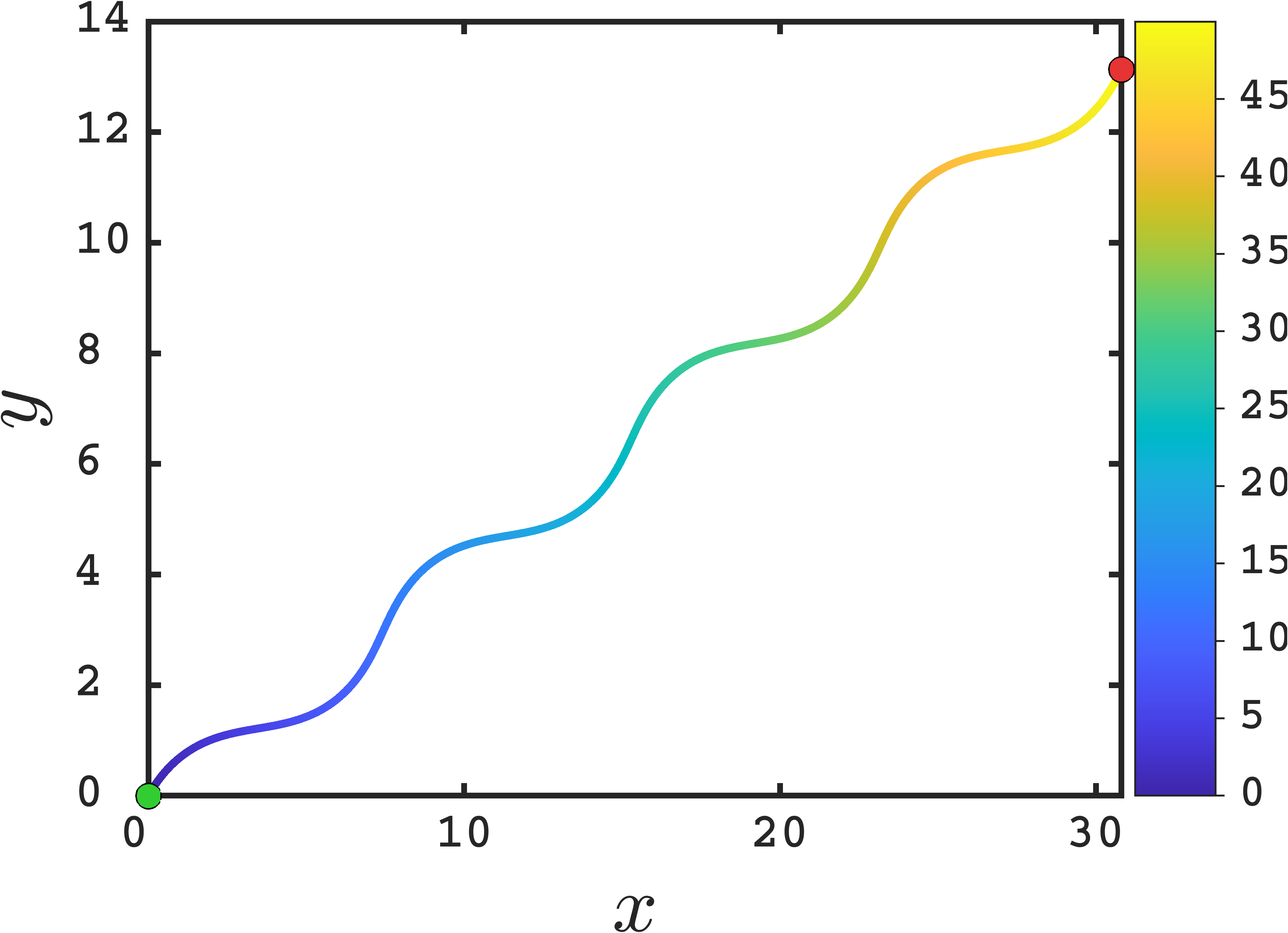}
        \caption{Particle's trajectory in field configuration case \RNum{1}.}
        \label{fig:fft-1a}
    \end{subfigure}
    \hfill
    \begin{subfigure}{0.45\textwidth}
        \centering
        \includegraphics[width=\linewidth]{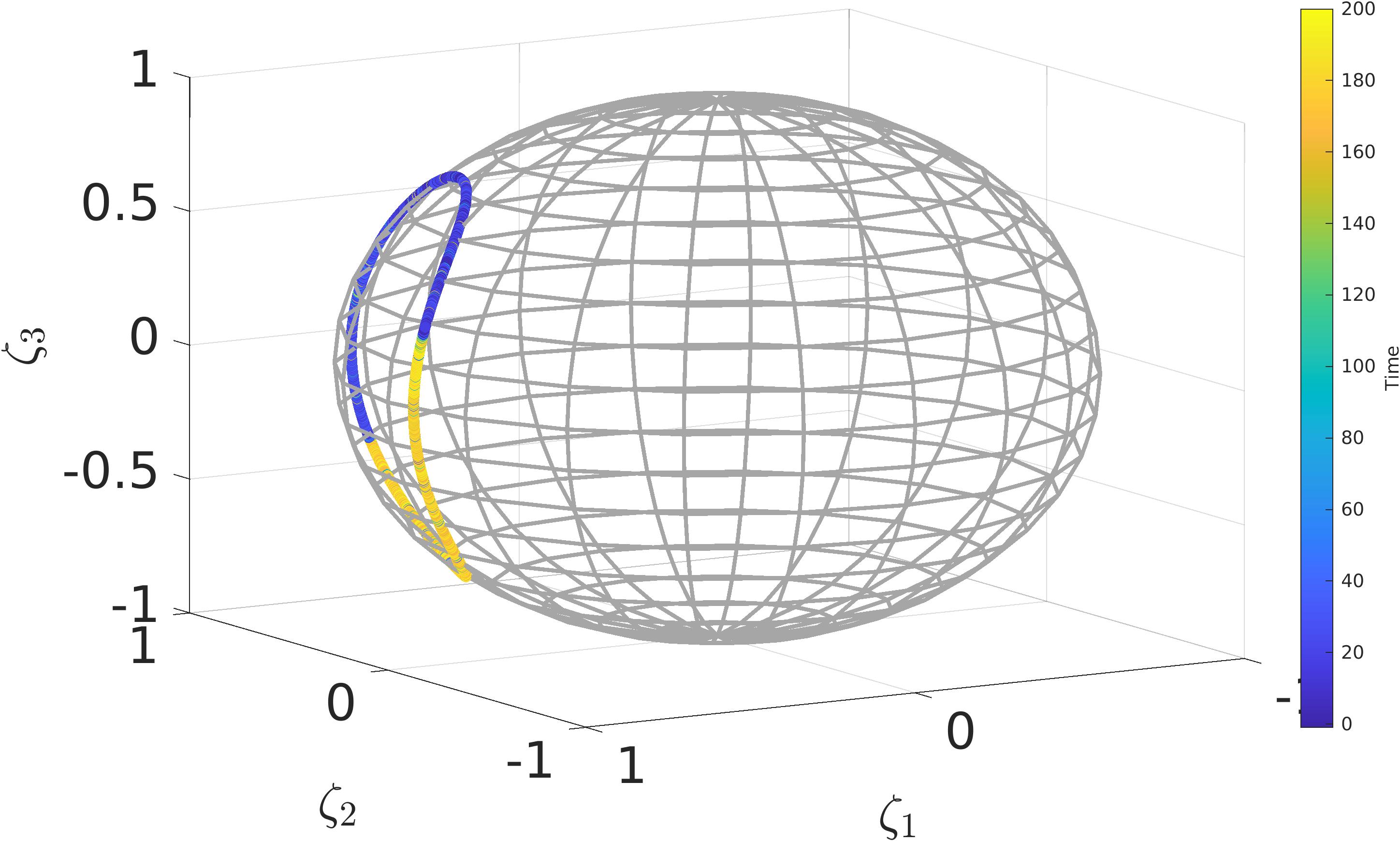}
        \caption{Evolution of the color charge in internal phase space}
        \label{fig:fft-1b}
    \end{subfigure}
    \caption{Particle trajectory (left panel) and corresponding color charge evolution (right panel) for field configuration: case \RNum{1} with $\alpha_y = 0.5$. Parameters are $\gamma = 1$ and $\Pi_x = \Pi_y = 0.61$. The color map represents time.}
    \label{fig:config1-dynamics}
\end{figure*}

\begin{figure*}[t]
    \centering
    \includegraphics[width=0.45\linewidth]{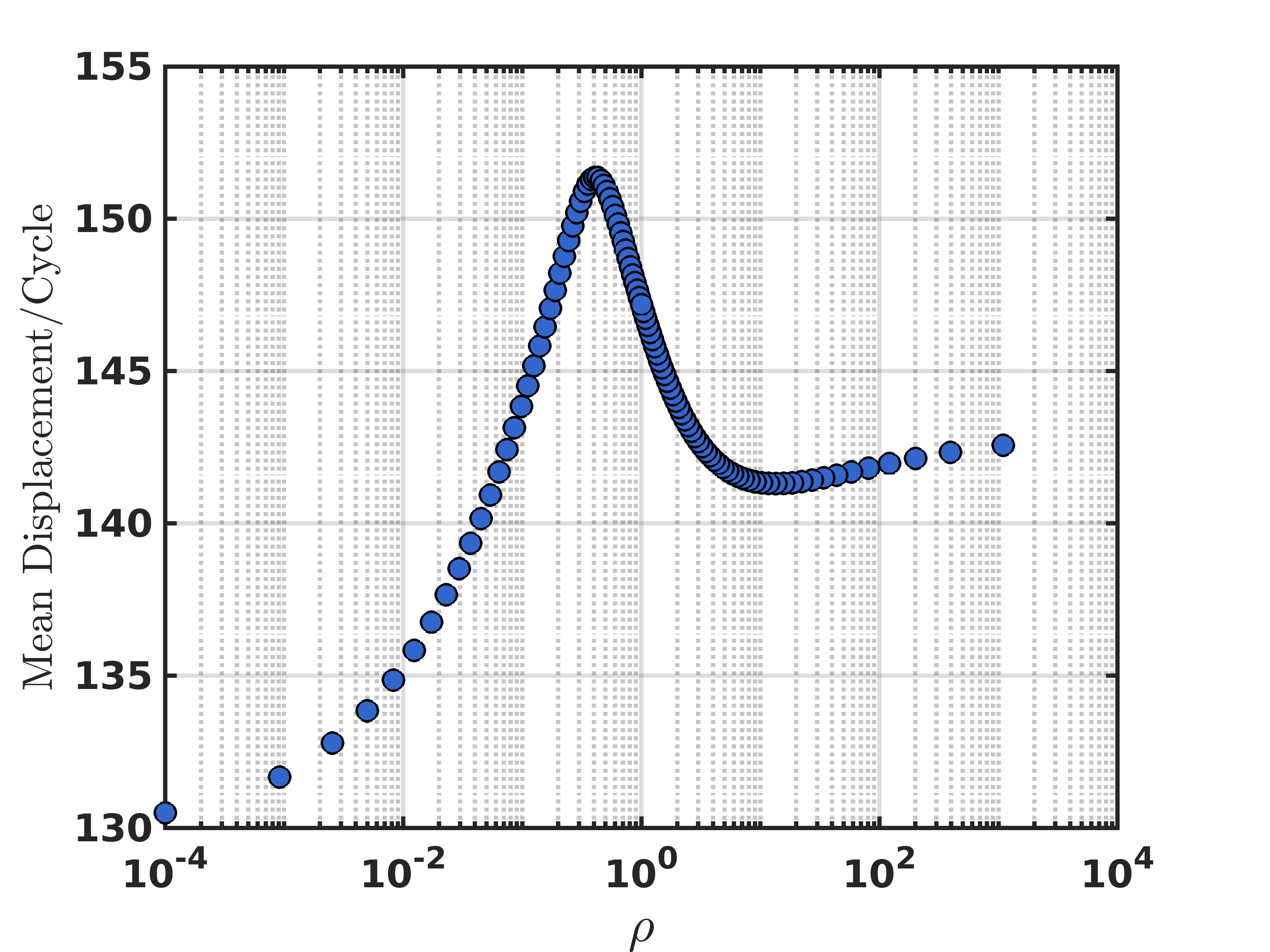}
    \caption{Illustrates the mean displacement per cycle as a function of $\rho$. The $x-$ axis of the plot is in $log$ scale. Here, initial conditions are $p(0)=1.0$ with $p_x(0)=p_y(0)$ and $\zeta_1(0)=\zeta_2(0) = \zeta_3(0)=0.57$. The value of $\alpha_y = 0.5$}
    \label{fig:cfg1-drift-vs-rho}
\end{figure*}

A representative particle trajectory illustrating a drift distinct from the Abelian case is shown in Fig.~\ref{fig:config1-dynamics}. The corresponding evolution of the color charge is also presented in the same figure. The drift is not along the $x-$ direction, as one might anticipate from the ED, but instead occurs in the $x-y$ plane. This can be understood by the fact that the magnetic field by itself induces a drift in the plane perpendicular to it. Further, this figure corresponds to $\alpha_y = 0.5$; as the value of the $\alpha_y$ increases, the drift along the $y-$ direction becomes dominant, although this is not shown here. In Fig. \ref{fig:cfg1-drift-vs-rho} we show how the mean displacement over a cycle varies with $\rho$. The illustrative figure corresponds to symmetric initial conditions and $\alpha_y = 0.5$. As $\rho$ (or $\rho^{-1}$) increases, the mean displacement per cycle tends to saturate. A similar trend is observed for other initial conditions and values of $\alpha_y$ which are not shown here. One can compare it with the Fig.\ref{fig: drift_one_comp} and quantitatively estimate the effect of color electric field.

Having established the dynamics in a configuration where the color electric and magnetic fields are aligned in internal space, we now turn to case \RNum{2}, in which the relative orientation of the electric and magnetic fields is changed.
\subsection{\label{sec:config2}Particle Motion in Field Configuration- Case \RNum{2}}
In this subsection, we analyse the test particle motion in the field configuration - case \RNum{2}, in which the scalar potential is chosen to be perpendicular, in internal space, to both vector potentials that generate the color magnetic field. The field configuration can be found in subsection \ref{subsec:config1} in equation \eqref{eq:fc-case2}. Here, the color magnetic field is in the $z$ direction in real space and along the $\hat{e}_3$ direction in the color space and the color electric field is present in $z-y$ plane in real space and in the color space lies in $\hat{e}_1-\hat{e}_2$ plane. The set of equations governing the dynamics can be obtained from the normalised version of equations in \eqref{eq:dyn}. The components of conserved canonical momentum and the total energy for this configuration are
\begin{eqnarray}
    \label{eq:pi-eb2}
    \Pi_x &=& p_x + \zeta_2 a_x, \qquad \Pi_y = p_y + \zeta_1 a_y, \nonumber \\ 
    \varepsilon &=& \gamma + \zeta_3 \varphi 
\end{eqnarray}

As the color fields do not exert any force along the $z$ direction, the corresponding momentum component remains constant and is set to zero without loss of generality. The momentum and energy continue to obey the bounds describes in equation 35


\begin{figure*}[t]
    \centering
    \begin{subfigure}{0.45\textwidth}
        \centering
        \includegraphics[width=\linewidth]{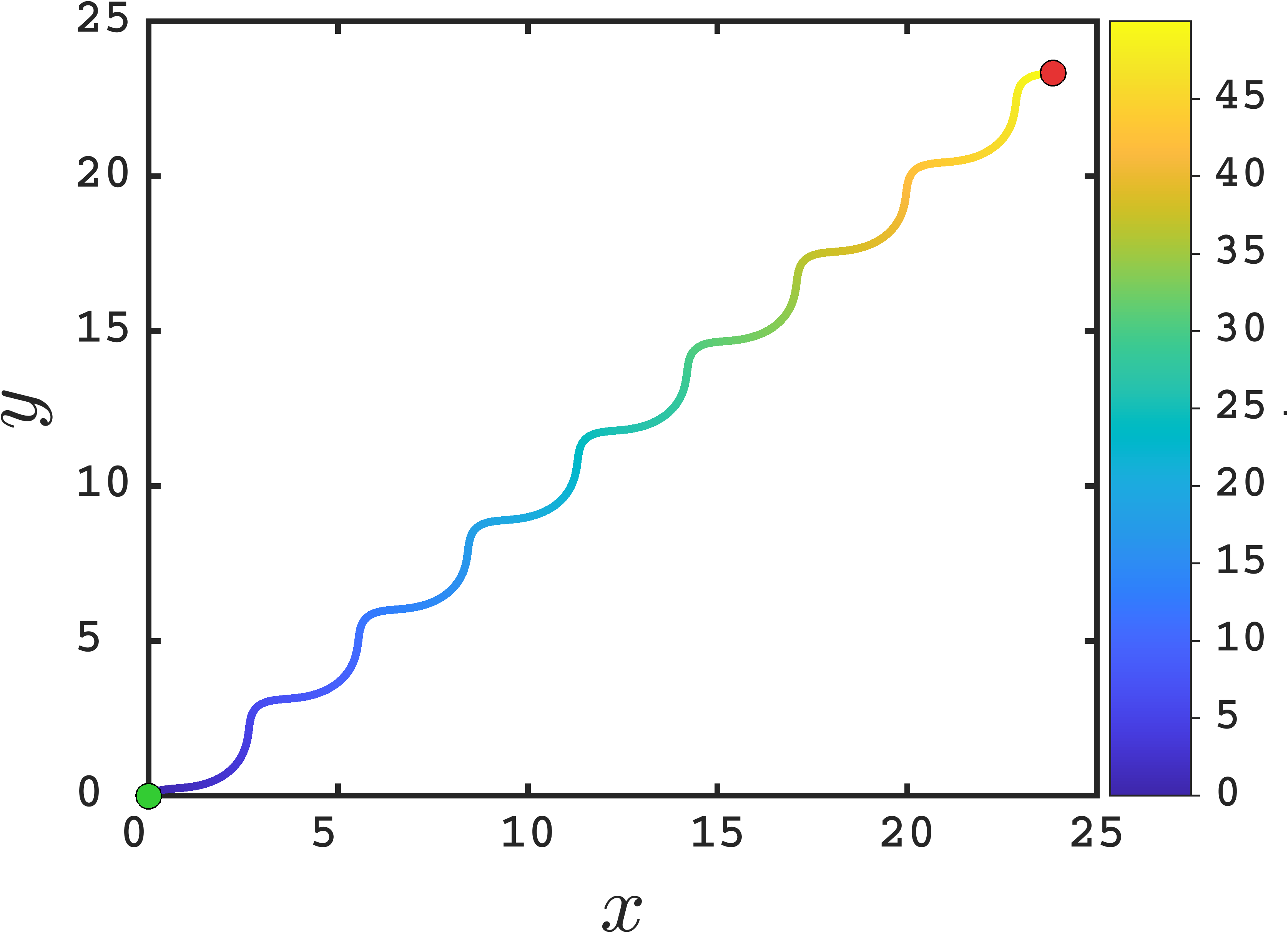}
        \caption{Particle's trajectory in field configuration case \RNum{2}.}
        \label{fig:fft-1a}
    \end{subfigure}
    \hfill
    \begin{subfigure}{0.45\textwidth}
        \centering
        \includegraphics[width=\linewidth]{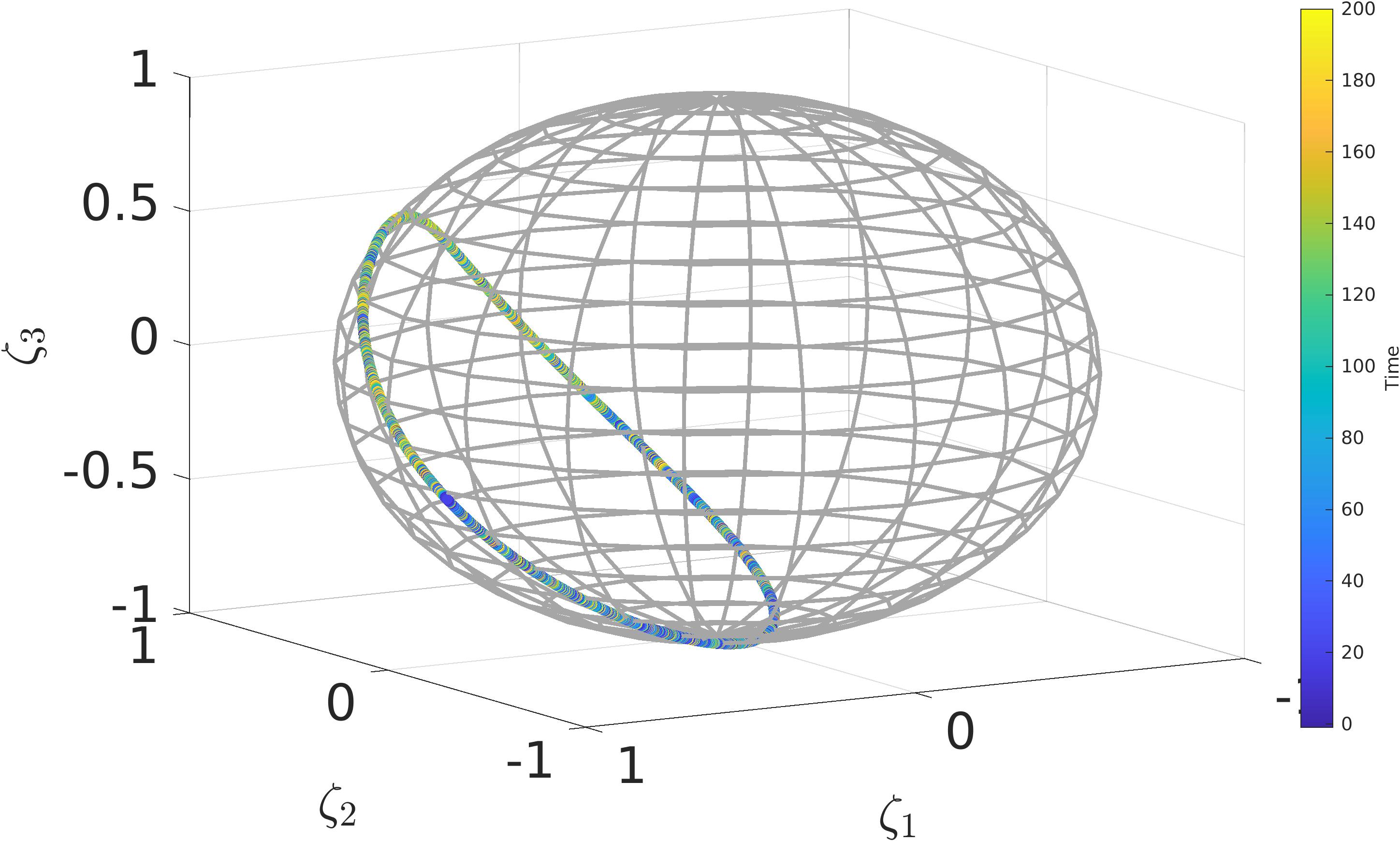}
        \caption{Evolution of the color charge in internal phase space}
        \label{fig:fft-1b}
    \end{subfigure}
    \caption{Particle trajectory (left panel) and corresponding color charge evolution (right panel) for field configuration: case \RNum{2} with $\alpha_x = 0.5$. Once we have chosen $a_x,a_y$ and $\alpha_x$, we cannot choose $\alpha_{y}$ independently. Parameters are $\gamma = 1$ and $\Pi_x = \Pi_y = 0.61$. The color map represents time.}
    \label{fig:config3_traj}
\end{figure*}

\begin{figure*}[t]
    \centering
    \includegraphics[width=0.45\linewidth]{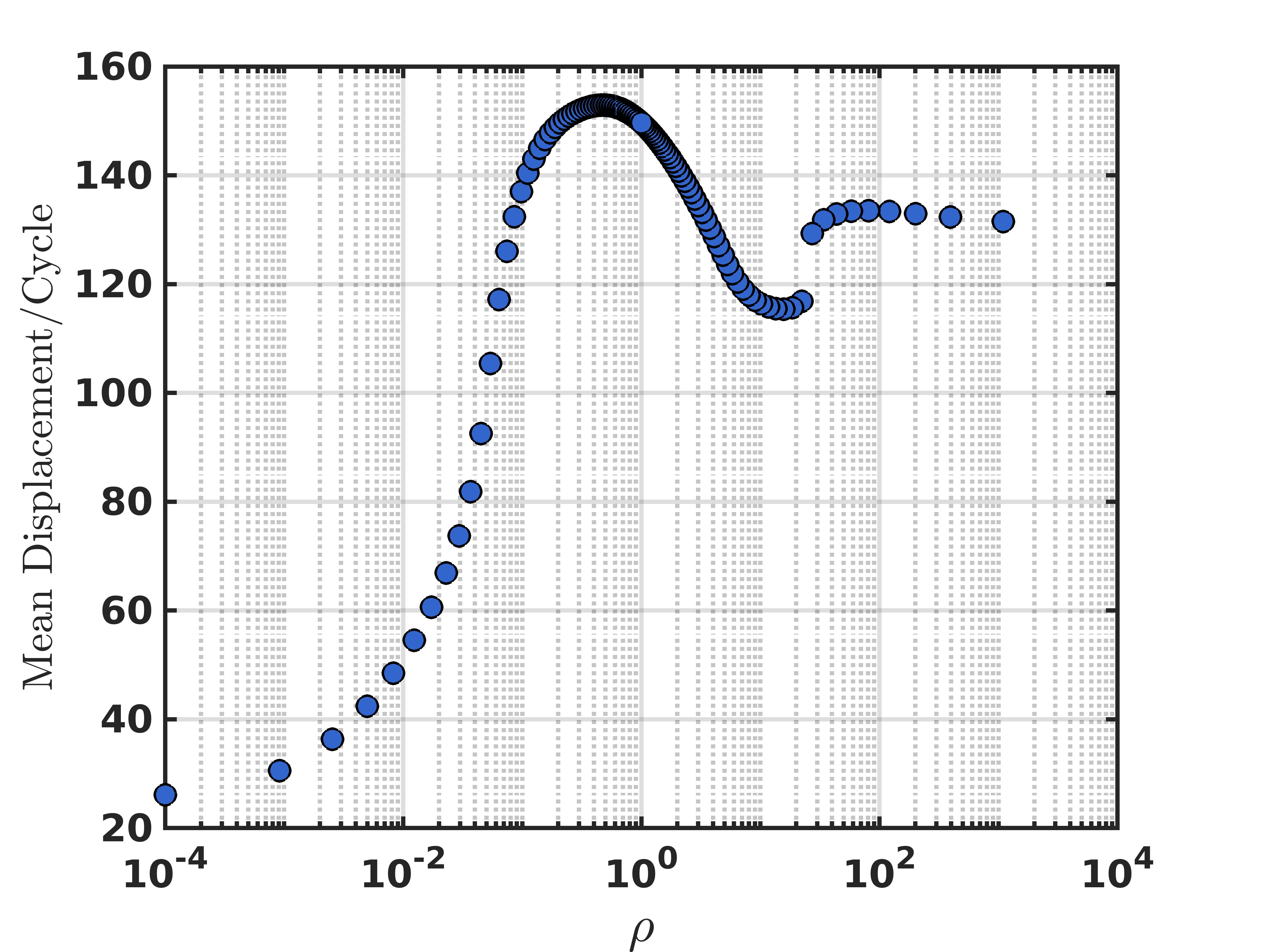}
    \caption{Illustrates the mean displacement per cycle as a function of $\rho$. The $x-$ axis of the plot is in $log$ scale. Here, initial conditions are $p(0)=1.0$ with $p_x(0)=p_y(0)$ and $\zeta_1(0)=\zeta_2(0) = \zeta_3(0)=0.57$. The value of $\alpha_x = 0.5$}
    \label{fig:cfg2-drift-vs-rho}
\end{figure*}

This field configuration produces a net drift in the $x$–$y$ plane. However, in contrast to the Abelian case (ED), the magnitude and direction of the drift are not determined solely by the ratio of the electric and magnetic field strengths. Moreover, unlike case \RNum{1}, no Lorentz transformation exists that can eliminate either the color electric field or the color magnetic field in a suitable frame, as they point in different internal directions. The equations of motion do not admit a closed-form analytical solution. We therefore solve the equations of motion numerically and present a representative particle trajectory exhibiting planar drift in Fig.~\ref{fig:config3_traj}. The corresponding evolution of the color charge can be found in the same figure. Further, in Fig. \ref{fig:cfg2-drift-vs-rho} we have shown the mean displacement over a period as a function of $\rho$ for $\alpha_{x} = 0.5$. The mean displacement tends to saturate as $\rho(\rho^{-1})$ increases. Similar trends are observed for other values of $\alpha_x$ and for different initial conditions, which are not shown here.

Taken together with the preceding configuration, these results demonstrate that particle motion in combined constant color electric and magnetic fields is qualitatively richer than in ED. Even at the level of test particle dynamics, YM interactions introduce drift that has no direct ED analogue.

\section{\label{sec:conclusions}Discussion and Conclusions}
In this work, we have carried out a systematic analysis of the motion of a test particle in constant color fields generated by static YM gauge potentials within the classical YM theory.  We first analysed particle dynamics in constant, maximally non-Abelian magnetic fields, identifying all admissible field configurations and examining the resulting trajectories. Our results show that, in contrast to the Abelian case, the motion of the test particle is unbounded in a one-component color magnetic field, while both bounded and unbounded trajectories arise in the three-component colour magnetic field. In the special case ($b_x = b_y = b_z$), the particle trajectory is bounded, and the cyclotron frequency is determined by an effective magnetic field that depends on the canonical momentum and the strength of the gauge potentials. The results presented in this work uunderscore the fact that the dynamics is governed by the underlying gauge potentials rather than the field strengths, highlighting the signature of non-Abelian dynamics. The effective magnetic field can be made zero with appropriate initial conditions such that the particle does not respond to the color magnetic field. Furthermore, we have shown that, in combined color electric and magnetic fields, the particle does not exhibit the conventional $E \times B$ drift characteristic of ED. 

The one-component color magnetic field is particularly relevant to systems with spin–orbit coupling, where effective non-Abelian gauge potentials encode the coupling between spin and momentum. From a spintronics perspective, our findings reveal how spin dynamics alone could induce nonlinear particle motion and drift, even in spatially uniform backgrounds and in the absence of electric fields. This suggests that semiclassical carrier trajectories in spin–orbit coupled materials need not form closed orbits. As mentioned, it may be reflected in the behaviour of transverse spin currents or spin-Hall responses. In  addition to disorder and defects, the internal spin dynamics by itself can act as a dynamical source of transport anomalies.

In cold-atom systems, where such gauge field configurations can be engineered, the results presented here describe the classical limit of neutral atom dynamics. These systems offer a good experimental control and therefore provide a promising route for realising and verifying the particle dynamics analysed in this work. Moreover, since cold-atom systems have been proposed as simulators of non-Abelian dynamics relevant to high-energy physics, our results may help connect single particle motion to bulk properties in systems such as the quark–gluon plasma (QGP) or in materials hosting synthetic non-Abelian gauge fields. An especially intriguing situation arises when the effective magnetic field discussed in Sec.~\ref{sec:Magdyn-3D} vanishes while the underlying gauge potentials remain non-zero. In this case, the medium becomes effectively transparent despite the presence of gauge fields, potentially leading to qualitatively different bulk behaviour.

Related synthetic gauge fields also appear in metamaterials and in photonic and optical systems, where they are used to control the polarisation and propagation of light. In such systems, the evolution of the light’s polarisation follows dynamics analogous to those described here. Because these systems can, in principle, host combinations of color electric and magnetic fields, they provide another potential arena for realising and testing the dynamics studied in this work, and for emulating aspects of high-energy gauge systems in laboratory environments \cite{zohar2013quantum,PRXQuantum.QCD}.

Finally, beyond condensed-matter, atomic, and photonic implementations, the configurations considered here are of interest from a high-energy perspective. Constant non-Abelian electric and magnetic fields play an important role in non-Abelian plasmas, including the quark–gluon plasma, where collective behaviour is governed by YM interactions. Recent studies have demonstrated the emergence of unstable modes arising purely from non-Abelian effects \cite{KNmine}. In this context, synthetic gauge-field systems may serve as experimentally accessible simulators of non-Abelian gauge dynamics \cite{rico2018so}, with the present test particle analysis providing a building block for connecting single particle motion to collective phenomena.

Taken together, our results establish constant non-Abelian gauge backgrounds as a minimal yet powerful platform for exploring dynamical effects that have no Abelian counterparts. Even at the level of classical test particle motion, YM interactions generate qualitatively new behaviour, underscoring the fundamental role of non-Abelian gauge potentials in the particle's dynamics across diverse physical systems.\\

\noindent\textbf{Acknowledgments:} The work of SB is supported by the Department of Science and Technology (DST), Govt. of India under DST/INSPIRE Fellowship/2019/IF190723. AD acknowledges support Anusandhan National Research Foundation (ANRF), Government of India, core grant CRG/2022/002782 as well as J C Bose Fellowship grant ANRF/JBG/2025/000237/PS.


\begin{thebibliography}{33}%
\makeatletter
\providecommand \@ifxundefined [1]{%
 \@ifx{#1\undefined}
}%
\providecommand \@ifnum [1]{%
 \ifnum #1\expandafter \@firstoftwo
 \else \expandafter \@secondoftwo
 \fi
}%
\providecommand \@ifx [1]{%
 \ifx #1\expandafter \@firstoftwo
 \else \expandafter \@secondoftwo
 \fi
}%
\providecommand \natexlab [1]{#1}%
\providecommand \enquote  [1]{``#1''}%
\providecommand \bibnamefont  [1]{#1}%
\providecommand \bibfnamefont [1]{#1}%
\providecommand \citenamefont [1]{#1}%
\providecommand \href@noop [0]{\@secondoftwo}%
\providecommand \href [0]{\begingroup \@sanitize@url \@href}%
\providecommand \@href[1]{\@@startlink{#1}\@@href}%
\providecommand \@@href[1]{\endgroup#1\@@endlink}%
\providecommand \@sanitize@url [0]{\catcode `\\12\catcode `\$12\catcode `\&12\catcode `\#12\catcode `\^12\catcode `\_12\catcode `\%12\relax}%
\providecommand \@@startlink[1]{}%
\providecommand \@@endlink[0]{}%
\providecommand \url  [0]{\begingroup\@sanitize@url \@url }%
\providecommand \@url [1]{\endgroup\@href {#1}{\urlprefix }}%
\providecommand \urlprefix  [0]{URL }%
\providecommand \Eprint [0]{\href }%
\providecommand \doibase [0]{https://doi.org/}%
\providecommand \selectlanguage [0]{\@gobble}%
\providecommand \bibinfo  [0]{\@secondoftwo}%
\providecommand \bibfield  [0]{\@secondoftwo}%
\providecommand \translation [1]{[#1]}%
\providecommand \BibitemOpen [0]{}%
\providecommand \bibitemStop [0]{}%
\providecommand \bibitemNoStop [0]{.\EOS\space}%
\providecommand \EOS [0]{\spacefactor3000\relax}%
\providecommand \BibitemShut  [1]{\csname bibitem#1\endcsname}%
\let\auto@bib@innerbib\@empty
\bibitem [{\citenamefont {Yang}\ and\ \citenamefont {Mills}(1954)}]{YangMills-original}%
  \BibitemOpen
  \bibfield  {author} {\bibinfo {author} {\bibfnamefont {C.~N.}\ \bibnamefont {Yang}}\ and\ \bibinfo {author} {\bibfnamefont {R.~L.}\ \bibnamefont {Mills}},\ }\bibfield  {title} {\bibinfo {title} {Conservation of isotopic spin and isotopic gauge invariance},\ }\href {https://doi.org/10.1103/PhysRev.96.191} {\bibfield  {journal} {\bibinfo  {journal} {Phys. Rev.}\ }\textbf {\bibinfo {volume} {96}},\ \bibinfo {pages} {191} (\bibinfo {year} {1954})}\BibitemShut {NoStop}%
\bibitem [{\citenamefont {Osterloh}\ \emph {et~al.}(2005)\citenamefont {Osterloh}, \citenamefont {Baig}, \citenamefont {Santos}, \citenamefont {Zoller},\ and\ \citenamefont {Lewenstein}}]{ColdAtomPRLOsterloh}%
  \BibitemOpen
  \bibfield  {author} {\bibinfo {author} {\bibfnamefont {K.}~\bibnamefont {Osterloh}}, \bibinfo {author} {\bibfnamefont {M.}~\bibnamefont {Baig}}, \bibinfo {author} {\bibfnamefont {L.}~\bibnamefont {Santos}}, \bibinfo {author} {\bibfnamefont {P.}~\bibnamefont {Zoller}},\ and\ \bibinfo {author} {\bibfnamefont {M.}~\bibnamefont {Lewenstein}},\ }\bibfield  {title} {\bibinfo {title} {Cold atoms in non-abelian gauge potentials: From the hofstadter "moth" to lattice gauge theory},\ }\href {https://doi.org/10.1103/PhysRevLett.95.010403} {\bibfield  {journal} {\bibinfo  {journal} {Phys. Rev. Lett.}\ }\textbf {\bibinfo {volume} {95}},\ \bibinfo {pages} {010403} (\bibinfo {year} {2005})}\BibitemShut {NoStop}%
\bibitem [{\citenamefont {Dalibard}\ \emph {et~al.}(2011)\citenamefont {Dalibard}, \citenamefont {Gerbier}, \citenamefont {Juzeli\ifmmode~\bar{u}\else \={u}\fi{}nas},\ and\ \citenamefont {\"Ohberg}}]{ColloquiumArtificialG}%
  \BibitemOpen
  \bibfield  {author} {\bibinfo {author} {\bibfnamefont {J.}~\bibnamefont {Dalibard}}, \bibinfo {author} {\bibfnamefont {F.}~\bibnamefont {Gerbier}}, \bibinfo {author} {\bibfnamefont {G.}~\bibnamefont {Juzeli\ifmmode~\bar{u}\else \={u}\fi{}nas}},\ and\ \bibinfo {author} {\bibfnamefont {P.}~\bibnamefont {\"Ohberg}},\ }\bibfield  {title} {\bibinfo {title} {Colloquium: Artificial gauge potentials for neutral atoms},\ }\href {https://doi.org/10.1103/RevModPhys.83.1523} {\bibfield  {journal} {\bibinfo  {journal} {Rev. Mod. Phys.}\ }\textbf {\bibinfo {volume} {83}},\ \bibinfo {pages} {1523} (\bibinfo {year} {2011})}\BibitemShut {NoStop}%
\bibitem [{\citenamefont {Yang}\ \emph {et~al.}(2019)\citenamefont {Yang}, \citenamefont {Peng}, \citenamefont {Zhu}, \citenamefont {Buljan}, \citenamefont {Joannopoulos}, \citenamefont {Zhen},\ and\ \citenamefont {Soljačić}}]{YiYangSyntAndObsv}%
  \BibitemOpen
  \bibfield  {author} {\bibinfo {author} {\bibfnamefont {Y.}~\bibnamefont {Yang}}, \bibinfo {author} {\bibfnamefont {C.}~\bibnamefont {Peng}}, \bibinfo {author} {\bibfnamefont {D.}~\bibnamefont {Zhu}}, \bibinfo {author} {\bibfnamefont {H.}~\bibnamefont {Buljan}}, \bibinfo {author} {\bibfnamefont {J.~D.}\ \bibnamefont {Joannopoulos}}, \bibinfo {author} {\bibfnamefont {B.}~\bibnamefont {Zhen}},\ and\ \bibinfo {author} {\bibfnamefont {M.}~\bibnamefont {Soljačić}},\ }\bibfield  {title} {\bibinfo {title} {Synthesis and observation of non-abelian gauge fields in real space},\ }\href {https://doi.org/10.1126/science.aay3183} {\bibfield  {journal} {\bibinfo  {journal} {Science}\ }\textbf {\bibinfo {volume} {365}},\ \bibinfo {pages} {1021} (\bibinfo {year} {2019})},\ \Eprint {https://arxiv.org/abs/https://www.science.org/doi/pdf/10.1126/science.aay3183} {https://www.science.org/doi/pdf/10.1126/science.aay3183} \BibitemShut {NoStop}%
\bibitem [{\citenamefont {Polimeno}\ \emph {et~al.}(2021)\citenamefont {Polimeno}, \citenamefont {Fieramosca}, \citenamefont {Lerario}, \citenamefont {Marco}, \citenamefont {Giorgi}, \citenamefont {Ballarini}, \citenamefont {Dominici}, \citenamefont {Ardizzone}, \citenamefont {Pugliese}, \citenamefont {Prontera}, \citenamefont {Maiorano}, \citenamefont {Gigli}, \citenamefont {Leblanc}, \citenamefont {Malpuech}, \citenamefont {Solnyshkov},\ and\ \citenamefont {Sanvitto}}]{PolimenoExptNAinPerovskite}%
  \BibitemOpen
  \bibfield  {author} {\bibinfo {author} {\bibfnamefont {L.}~\bibnamefont {Polimeno}}, \bibinfo {author} {\bibfnamefont {A.}~\bibnamefont {Fieramosca}}, \bibinfo {author} {\bibfnamefont {G.}~\bibnamefont {Lerario}}, \bibinfo {author} {\bibfnamefont {L.~D.}\ \bibnamefont {Marco}}, \bibinfo {author} {\bibfnamefont {M.~D.}\ \bibnamefont {Giorgi}}, \bibinfo {author} {\bibfnamefont {D.}~\bibnamefont {Ballarini}}, \bibinfo {author} {\bibfnamefont {L.}~\bibnamefont {Dominici}}, \bibinfo {author} {\bibfnamefont {V.}~\bibnamefont {Ardizzone}}, \bibinfo {author} {\bibfnamefont {M.}~\bibnamefont {Pugliese}}, \bibinfo {author} {\bibfnamefont {C.~T.}\ \bibnamefont {Prontera}}, \bibinfo {author} {\bibfnamefont {V.}~\bibnamefont {Maiorano}}, \bibinfo {author} {\bibfnamefont {G.}~\bibnamefont {Gigli}}, \bibinfo {author} {\bibfnamefont {C.}~\bibnamefont {Leblanc}}, \bibinfo {author} {\bibfnamefont {G.}~\bibnamefont {Malpuech}}, \bibinfo {author} {\bibfnamefont {D.~D.}\ \bibnamefont {Solnyshkov}},\ and\ \bibinfo {author}
  {\bibfnamefont {D.}~\bibnamefont {Sanvitto}},\ }\bibfield  {title} {\bibinfo {title} {Experimental investigation of a non-abelian gauge field in 2d perovskite photonic platform},\ }\href {https://doi.org/10.1364/OPTICA.427088} {\bibfield  {journal} {\bibinfo  {journal} {Optica}\ }\textbf {\bibinfo {volume} {8}},\ \bibinfo {pages} {1442} (\bibinfo {year} {2021})}\BibitemShut {NoStop}%
\bibitem [{\citenamefont {Yang}\ \emph {et~al.}(2024)\citenamefont {Yang}, \citenamefont {Yang}, \citenamefont {Ma}, \citenamefont {Li}, \citenamefont {Zhang},\ and\ \citenamefont {Chan}}]{YiYangNAinLightandSound}%
  \BibitemOpen
  \bibfield  {author} {\bibinfo {author} {\bibfnamefont {Y.}~\bibnamefont {Yang}}, \bibinfo {author} {\bibfnamefont {B.}~\bibnamefont {Yang}}, \bibinfo {author} {\bibfnamefont {G.}~\bibnamefont {Ma}}, \bibinfo {author} {\bibfnamefont {J.}~\bibnamefont {Li}}, \bibinfo {author} {\bibfnamefont {S.}~\bibnamefont {Zhang}},\ and\ \bibinfo {author} {\bibfnamefont {C.~T.}\ \bibnamefont {Chan}},\ }\bibfield  {title} {\bibinfo {title} {Non-abelian physics in light and sound},\ }\href {https://doi.org/10.1126/science.adf9621} {\bibfield  {journal} {\bibinfo  {journal} {Science}\ }\textbf {\bibinfo {volume} {383}},\ \bibinfo {pages} {eadf9621} (\bibinfo {year} {2024})},\ \Eprint {https://arxiv.org/abs/https://www.science.org/doi/pdf/10.1126/science.adf9621} {https://www.science.org/doi/pdf/10.1126/science.adf9621} \BibitemShut {NoStop}%
\bibitem [{\citenamefont {Tan}\ \emph {et~al.}(2020)\citenamefont {Tan}, \citenamefont {Chen}, \citenamefont {Ho}, \citenamefont {Huang}, \citenamefont {Jalil}, \citenamefont {Chang},\ and\ \citenamefont {Murakami}}]{YMSpinTAN}%
  \BibitemOpen
  \bibfield  {author} {\bibinfo {author} {\bibfnamefont {S.~G.}\ \bibnamefont {Tan}}, \bibinfo {author} {\bibfnamefont {S.-H.}\ \bibnamefont {Chen}}, \bibinfo {author} {\bibfnamefont {C.~S.}\ \bibnamefont {Ho}}, \bibinfo {author} {\bibfnamefont {C.-C.}\ \bibnamefont {Huang}}, \bibinfo {author} {\bibfnamefont {M.~B.}\ \bibnamefont {Jalil}}, \bibinfo {author} {\bibfnamefont {C.~R.}\ \bibnamefont {Chang}},\ and\ \bibinfo {author} {\bibfnamefont {S.}~\bibnamefont {Murakami}},\ }\bibfield  {title} {\bibinfo {title} {Yang–mills physics in spintronics},\ }\href {https://doi.org/https://doi.org/10.1016/j.physrep.2020.08.002} {\bibfield  {journal} {\bibinfo  {journal} {Physics Reports}\ }\textbf {\bibinfo {volume} {882}},\ \bibinfo {pages} {1} (\bibinfo {year} {2020})},\ \bibinfo {note} {yang-Mills physics in spintronics}\BibitemShut {NoStop}%
\bibitem [{\citenamefont {Hao}\ and\ \citenamefont {Xiao}(2015)}]{PhysRevApplied.SHEswitch}%
  \BibitemOpen
  \bibfield  {author} {\bibinfo {author} {\bibfnamefont {Q.}~\bibnamefont {Hao}}\ and\ \bibinfo {author} {\bibfnamefont {G.}~\bibnamefont {Xiao}},\ }\bibfield  {title} {\bibinfo {title} {Giant spin hall effect and switching induced by spin-transfer torque in a $\mathrm{W}/{\mathrm{co}}_{40}{\mathrm{fe}}_{40}{\mathrm{b}}_{20}/\mathrm{MgO}$ structure with perpendicular magnetic anisotropy},\ }\href {https://doi.org/10.1103/PhysRevApplied.3.034009} {\bibfield  {journal} {\bibinfo  {journal} {Phys. Rev. Appl.}\ }\textbf {\bibinfo {volume} {3}},\ \bibinfo {pages} {034009} (\bibinfo {year} {2015})}\BibitemShut {NoStop}%
\bibitem [{\citenamefont {Hu}\ \emph {et~al.}(2022)\citenamefont {Hu}, \citenamefont {Shao}, \citenamefont {Yang}, \citenamefont {Pan}, \citenamefont {Fu}, \citenamefont {Tang}, \citenamefont {Yang}, \citenamefont {Fan}, \citenamefont {Zhou}, \citenamefont {Tsymbal},\ and\ \citenamefont {Qiu}}]{SHE.switch}%
  \BibitemOpen
  \bibfield  {author} {\bibinfo {author} {\bibfnamefont {S.}~\bibnamefont {Hu}}, \bibinfo {author} {\bibfnamefont {D.-F.}\ \bibnamefont {Shao}}, \bibinfo {author} {\bibfnamefont {H.}~\bibnamefont {Yang}}, \bibinfo {author} {\bibfnamefont {C.}~\bibnamefont {Pan}}, \bibinfo {author} {\bibfnamefont {Z.}~\bibnamefont {Fu}}, \bibinfo {author} {\bibfnamefont {M.}~\bibnamefont {Tang}}, \bibinfo {author} {\bibfnamefont {Y.}~\bibnamefont {Yang}}, \bibinfo {author} {\bibfnamefont {W.}~\bibnamefont {Fan}}, \bibinfo {author} {\bibfnamefont {S.}~\bibnamefont {Zhou}}, \bibinfo {author} {\bibfnamefont {E.~Y.}\ \bibnamefont {Tsymbal}},\ and\ \bibinfo {author} {\bibfnamefont {X.}~\bibnamefont {Qiu}},\ }\bibfield  {title} {\bibinfo {title} {Efficient perpendicular magnetization switching by a magnetic spin hall effect in a noncollinear antiferromagnet},\ }\href {https://doi.org/10.1038/s41467-022-32179-2} {\bibfield  {journal} {\bibinfo  {journal} {Nature Communications}\ }\textbf {\bibinfo {volume} {13}},\ \bibinfo {pages}
  {4447} (\bibinfo {year} {2022})}\BibitemShut {NoStop}%
\bibitem [{\citenamefont {Sheng}\ \emph {et~al.}(2023)\citenamefont {Sheng}, \citenamefont {Chen}, \citenamefont {Yuan}, \citenamefont {Liu}, \citenamefont {Zhang}, \citenamefont {Jing}, \citenamefont {Kuang},\ and\ \citenamefont {Zhou}}]{PSHE.PQE}%
  \BibitemOpen
  \bibfield  {author} {\bibinfo {author} {\bibfnamefont {L.}~\bibnamefont {Sheng}}, \bibinfo {author} {\bibfnamefont {Y.}~\bibnamefont {Chen}}, \bibinfo {author} {\bibfnamefont {S.}~\bibnamefont {Yuan}}, \bibinfo {author} {\bibfnamefont {X.}~\bibnamefont {Liu}}, \bibinfo {author} {\bibfnamefont {Z.}~\bibnamefont {Zhang}}, \bibinfo {author} {\bibfnamefont {H.}~\bibnamefont {Jing}}, \bibinfo {author} {\bibfnamefont {L.-M.}\ \bibnamefont {Kuang}},\ and\ \bibinfo {author} {\bibfnamefont {X.}~\bibnamefont {Zhou}},\ }\bibfield  {title} {\bibinfo {title} {Photonic spin hall effect: Physics, manipulations, and applications},\ }\href {https://doi.org/https://doi.org/10.1016/j.pquantelec.2023.100484} {\bibfield  {journal} {\bibinfo  {journal} {Progress in Quantum Electronics}\ }\textbf {\bibinfo {volume} {91-92}},\ \bibinfo {pages} {100484} (\bibinfo {year} {2023})}\BibitemShut {NoStop}%
\bibitem [{\citenamefont {Cho}(2005)}]{chosuper}%
  \BibitemOpen
  \bibfield  {author} {\bibinfo {author} {\bibfnamefont {Y.~M.}\ \bibnamefont {Cho}},\ }\bibfield  {title} {\bibinfo {title} {Non-abelian superconductivity: An effective-field theory for two-gap superconductivity},\ }\href {https://doi.org/10.1103/PhysRevB.72.212516} {\bibfield  {journal} {\bibinfo  {journal} {Phys. Rev. B}\ }\textbf {\bibinfo {volume} {72}},\ \bibinfo {pages} {212516} (\bibinfo {year} {2005})}\BibitemShut {NoStop}%
\bibitem [{\citenamefont {Nitta}\ \emph {et~al.}(1997)\citenamefont {Nitta}, \citenamefont {Akazaki}, \citenamefont {Takayanagi},\ and\ \citenamefont {Enoki}}]{Nitta1997}%
  \BibitemOpen
  \bibfield  {author} {\bibinfo {author} {\bibfnamefont {J.}~\bibnamefont {Nitta}}, \bibinfo {author} {\bibfnamefont {T.}~\bibnamefont {Akazaki}}, \bibinfo {author} {\bibfnamefont {H.}~\bibnamefont {Takayanagi}},\ and\ \bibinfo {author} {\bibfnamefont {T.}~\bibnamefont {Enoki}},\ }\bibfield  {title} {\bibinfo {title} {Gate control of spin-orbit interaction in an inverted i n 0.53 g a 0.47 as/i n 0.52 a l 0.48 as heterostructure},\ }\href@noop {} {\bibfield  {journal} {\bibinfo  {journal} {Physical Review Letters}\ }\textbf {\bibinfo {volume} {78}},\ \bibinfo {pages} {1335} (\bibinfo {year} {1997})}\BibitemShut {NoStop}%
\bibitem [{\citenamefont {Grundler}(2000)}]{grundler2000large}%
  \BibitemOpen
  \bibfield  {author} {\bibinfo {author} {\bibfnamefont {D.}~\bibnamefont {Grundler}},\ }\bibfield  {title} {\bibinfo {title} {Large rashba splitting in inas quantum wells due to electron wave function penetration into the barrier layers},\ }\href@noop {} {\bibfield  {journal} {\bibinfo  {journal} {Physical review letters}\ }\textbf {\bibinfo {volume} {84}},\ \bibinfo {pages} {6074} (\bibinfo {year} {2000})}\BibitemShut {NoStop}%
\bibitem [{\citenamefont {Caviglia}\ \emph {et~al.}(2010)\citenamefont {Caviglia}, \citenamefont {Gabay}, \citenamefont {Gariglio}, \citenamefont {Reyren}, \citenamefont {Cancellieri},\ and\ \citenamefont {Triscone}}]{caviglia2010tunable}%
  \BibitemOpen
  \bibfield  {author} {\bibinfo {author} {\bibfnamefont {A.}~\bibnamefont {Caviglia}}, \bibinfo {author} {\bibfnamefont {M.}~\bibnamefont {Gabay}}, \bibinfo {author} {\bibfnamefont {S.}~\bibnamefont {Gariglio}}, \bibinfo {author} {\bibfnamefont {N.}~\bibnamefont {Reyren}}, \bibinfo {author} {\bibfnamefont {C.}~\bibnamefont {Cancellieri}},\ and\ \bibinfo {author} {\bibfnamefont {J.-M.}\ \bibnamefont {Triscone}},\ }\bibfield  {title} {\bibinfo {title} {Tunable rashba spin-orbit interaction at oxide interfaces},\ }\href@noop {} {\bibfield  {journal} {\bibinfo  {journal} {Physical review letters}\ }\textbf {\bibinfo {volume} {104}},\ \bibinfo {pages} {126803} (\bibinfo {year} {2010})}\BibitemShut {NoStop}%
\bibitem [{\citenamefont {Ruseckas}\ \emph {et~al.}(2005)\citenamefont {Ruseckas}, \citenamefont {Juzeli\ifmmode~\bar{u}\else \={u}\fi{}nas}, \citenamefont {\"Ohberg},\ and\ \citenamefont {Fleischhauer}}]{PRL.GenMag}%
  \BibitemOpen
  \bibfield  {author} {\bibinfo {author} {\bibfnamefont {J.}~\bibnamefont {Ruseckas}}, \bibinfo {author} {\bibfnamefont {G.}~\bibnamefont {Juzeli\ifmmode~\bar{u}\else \={u}\fi{}nas}}, \bibinfo {author} {\bibfnamefont {P.}~\bibnamefont {\"Ohberg}},\ and\ \bibinfo {author} {\bibfnamefont {M.}~\bibnamefont {Fleischhauer}},\ }\bibfield  {title} {\bibinfo {title} {Non-abelian gauge potentials for ultracold atoms with degenerate dark states},\ }\href {https://doi.org/10.1103/PhysRevLett.95.010404} {\bibfield  {journal} {\bibinfo  {journal} {Phys. Rev. Lett.}\ }\textbf {\bibinfo {volume} {95}},\ \bibinfo {pages} {010404} (\bibinfo {year} {2005})}\BibitemShut {NoStop}%
\bibitem [{\citenamefont {Rico}\ \emph {et~al.}(2018)\citenamefont {Rico}, \citenamefont {Dalmonte}, \citenamefont {Zoller}, \citenamefont {Banerjee}, \citenamefont {B{\"o}gli}, \citenamefont {Stebler},\ and\ \citenamefont {Wiese}}]{rico2018so}%
  \BibitemOpen
  \bibfield  {author} {\bibinfo {author} {\bibfnamefont {E.}~\bibnamefont {Rico}}, \bibinfo {author} {\bibfnamefont {M.}~\bibnamefont {Dalmonte}}, \bibinfo {author} {\bibfnamefont {P.}~\bibnamefont {Zoller}}, \bibinfo {author} {\bibfnamefont {D.}~\bibnamefont {Banerjee}}, \bibinfo {author} {\bibfnamefont {M.}~\bibnamefont {B{\"o}gli}}, \bibinfo {author} {\bibfnamefont {P.}~\bibnamefont {Stebler}},\ and\ \bibinfo {author} {\bibfnamefont {U.-J.}\ \bibnamefont {Wiese}},\ }\bibfield  {title} {\bibinfo {title} {So (3)“nuclear physics” with ultracold gases},\ }\href@noop {} {\bibfield  {journal} {\bibinfo  {journal} {Annals of physics}\ }\textbf {\bibinfo {volume} {393}},\ \bibinfo {pages} {466} (\bibinfo {year} {2018})}\BibitemShut {NoStop}%
\bibitem [{\citenamefont {Anderson}\ \emph {et~al.}(2012)\citenamefont {Anderson}, \citenamefont {Juzeli{\=u}nas}, \citenamefont {Galitski},\ and\ \citenamefont {Spielman}}]{anderson2012synthetic}%
  \BibitemOpen
  \bibfield  {author} {\bibinfo {author} {\bibfnamefont {B.~M.}\ \bibnamefont {Anderson}}, \bibinfo {author} {\bibfnamefont {G.}~\bibnamefont {Juzeli{\=u}nas}}, \bibinfo {author} {\bibfnamefont {V.~M.}\ \bibnamefont {Galitski}},\ and\ \bibinfo {author} {\bibfnamefont {I.~B.}\ \bibnamefont {Spielman}},\ }\bibfield  {title} {\bibinfo {title} {Synthetic 3d spin-orbit coupling},\ }\href@noop {} {\bibfield  {journal} {\bibinfo  {journal} {Physical review letters}\ }\textbf {\bibinfo {volume} {108}},\ \bibinfo {pages} {235301} (\bibinfo {year} {2012})}\BibitemShut {NoStop}%
\bibitem [{\citenamefont {Anderson}\ \emph {et~al.}(2013)\citenamefont {Anderson}, \citenamefont {Spielman},\ and\ \citenamefont {Juzeli{\=u}nas}}]{anderson2013magnetically}%
  \BibitemOpen
  \bibfield  {author} {\bibinfo {author} {\bibfnamefont {B.~M.}\ \bibnamefont {Anderson}}, \bibinfo {author} {\bibfnamefont {I.~B.}\ \bibnamefont {Spielman}},\ and\ \bibinfo {author} {\bibfnamefont {G.}~\bibnamefont {Juzeli{\=u}nas}},\ }\bibfield  {title} {\bibinfo {title} {Magnetically generated spin-orbit coupling for ultracold atoms},\ }\href@noop {} {\bibfield  {journal} {\bibinfo  {journal} {Physical review letters}\ }\textbf {\bibinfo {volume} {111}},\ \bibinfo {pages} {125301} (\bibinfo {year} {2013})}\BibitemShut {NoStop}%
\bibitem [{\citenamefont {Madasu}\ \emph {et~al.}(2025)\citenamefont {Madasu}, \citenamefont {Mitra}, \citenamefont {Gabardos}, \citenamefont {Rathod}, \citenamefont {Zanon-Willette}, \citenamefont {Miniatura}, \citenamefont {Chevy}, \citenamefont {Kwong},\ and\ \citenamefont {Wilkowski}}]{madasu2025experimental}%
  \BibitemOpen
  \bibfield  {author} {\bibinfo {author} {\bibfnamefont {C.~S.}\ \bibnamefont {Madasu}}, \bibinfo {author} {\bibfnamefont {C.}~\bibnamefont {Mitra}}, \bibinfo {author} {\bibfnamefont {L.}~\bibnamefont {Gabardos}}, \bibinfo {author} {\bibfnamefont {K.~D.}\ \bibnamefont {Rathod}}, \bibinfo {author} {\bibfnamefont {T.}~\bibnamefont {Zanon-Willette}}, \bibinfo {author} {\bibfnamefont {C.}~\bibnamefont {Miniatura}}, \bibinfo {author} {\bibfnamefont {F.}~\bibnamefont {Chevy}}, \bibinfo {author} {\bibfnamefont {C.~C.}\ \bibnamefont {Kwong}},\ and\ \bibinfo {author} {\bibfnamefont {D.}~\bibnamefont {Wilkowski}},\ }\bibfield  {title} {\bibinfo {title} {Experimental realization of a su (3) color-orbit coupling in an ultracold gas},\ }\href@noop {} {\bibfield  {journal} {\bibinfo  {journal} {Nature Communications}\ }\textbf {\bibinfo {volume} {16}},\ \bibinfo {pages} {8448} (\bibinfo {year} {2025})}\BibitemShut {NoStop}%
\bibitem [{\citenamefont {Liang}\ \emph {et~al.}(2024)\citenamefont {Liang}, \citenamefont {Dong}, \citenamefont {Pan}, \citenamefont {Wang}, \citenamefont {Li}, \citenamefont {Yang}, \citenamefont {Yi},\ and\ \citenamefont {Yan}}]{liang2024chiral}%
  \BibitemOpen
  \bibfield  {author} {\bibinfo {author} {\bibfnamefont {Q.}~\bibnamefont {Liang}}, \bibinfo {author} {\bibfnamefont {Z.}~\bibnamefont {Dong}}, \bibinfo {author} {\bibfnamefont {J.-S.}\ \bibnamefont {Pan}}, \bibinfo {author} {\bibfnamefont {H.}~\bibnamefont {Wang}}, \bibinfo {author} {\bibfnamefont {H.}~\bibnamefont {Li}}, \bibinfo {author} {\bibfnamefont {Z.}~\bibnamefont {Yang}}, \bibinfo {author} {\bibfnamefont {W.}~\bibnamefont {Yi}},\ and\ \bibinfo {author} {\bibfnamefont {B.}~\bibnamefont {Yan}},\ }\bibfield  {title} {\bibinfo {title} {Chiral dynamics of ultracold atoms under a tunable su (2) synthetic gauge field},\ }\href@noop {} {\bibfield  {journal} {\bibinfo  {journal} {Nature Physics}\ }\textbf {\bibinfo {volume} {20}},\ \bibinfo {pages} {1738} (\bibinfo {year} {2024})}\BibitemShut {NoStop}%
\bibitem [{\citenamefont {Liu}\ and\ \citenamefont {Li}(2015)}]{liu2015gauge}%
  \BibitemOpen
  \bibfield  {author} {\bibinfo {author} {\bibfnamefont {F.}~\bibnamefont {Liu}}\ and\ \bibinfo {author} {\bibfnamefont {J.}~\bibnamefont {Li}},\ }\bibfield  {title} {\bibinfo {title} {Gauge field optics with anisotropic media},\ }\href@noop {} {\bibfield  {journal} {\bibinfo  {journal} {Physical Review Letters}\ }\textbf {\bibinfo {volume} {114}},\ \bibinfo {pages} {103902} (\bibinfo {year} {2015})}\BibitemShut {NoStop}%
\bibitem [{\citenamefont {Chen}\ \emph {et~al.}(2019)\citenamefont {Chen}, \citenamefont {Zhang}, \citenamefont {Xiong}, \citenamefont {Hang}, \citenamefont {Li}, \citenamefont {Shen},\ and\ \citenamefont {Chan}}]{chen2019non}%
  \BibitemOpen
  \bibfield  {author} {\bibinfo {author} {\bibfnamefont {Y.}~\bibnamefont {Chen}}, \bibinfo {author} {\bibfnamefont {R.-Y.}\ \bibnamefont {Zhang}}, \bibinfo {author} {\bibfnamefont {Z.}~\bibnamefont {Xiong}}, \bibinfo {author} {\bibfnamefont {Z.~H.}\ \bibnamefont {Hang}}, \bibinfo {author} {\bibfnamefont {J.}~\bibnamefont {Li}}, \bibinfo {author} {\bibfnamefont {J.~Q.}\ \bibnamefont {Shen}},\ and\ \bibinfo {author} {\bibfnamefont {C.}~\bibnamefont {Chan}},\ }\bibfield  {title} {\bibinfo {title} {Non-abelian gauge field optics},\ }\href@noop {} {\bibfield  {journal} {\bibinfo  {journal} {Nature communications}\ }\textbf {\bibinfo {volume} {10}},\ \bibinfo {pages} {3125} (\bibinfo {year} {2019})}\BibitemShut {NoStop}%
\bibitem [{\citenamefont {Liu}\ \emph {et~al.}(2025)\citenamefont {Liu}, \citenamefont {Xu},\ and\ \citenamefont {Hang}}]{liu2025general}%
  \BibitemOpen
  \bibfield  {author} {\bibinfo {author} {\bibfnamefont {B.}~\bibnamefont {Liu}}, \bibinfo {author} {\bibfnamefont {T.}~\bibnamefont {Xu}},\ and\ \bibinfo {author} {\bibfnamefont {Z.~H.}\ \bibnamefont {Hang}},\ }\bibfield  {title} {\bibinfo {title} {A general recipe to observe non-abelian gauge field in metamaterials},\ }\href@noop {} {\bibfield  {journal} {\bibinfo  {journal} {Nanophotonics}\ }\textbf {\bibinfo {volume} {14}},\ \bibinfo {pages} {1135} (\bibinfo {year} {2025})}\BibitemShut {NoStop}%
\bibitem [{\citenamefont {Bhat}\ \emph {et~al.}(2025)\citenamefont {Bhat}, \citenamefont {Das}, \citenamefont {Paradkar},\ and\ \citenamefont {Ravishankar}}]{KN_2025}%
  \BibitemOpen
  \bibfield  {author} {\bibinfo {author} {\bibfnamefont {S.}~\bibnamefont {Bhat}}, \bibinfo {author} {\bibfnamefont {A.}~\bibnamefont {Das}}, \bibinfo {author} {\bibfnamefont {B.}~\bibnamefont {Paradkar}},\ and\ \bibinfo {author} {\bibfnamefont {V.}~\bibnamefont {Ravishankar}},\ }\bibfield  {title} {\bibinfo {title} {Experimental signatures for identifying distinct origins of color field generation},\ }\href {https://doi.org/10.1088/1367-2630/ae0820} {\bibfield  {journal} {\bibinfo  {journal} {New Journal of Physics}\ }\textbf {\bibinfo {volume} {27}},\ \bibinfo {pages} {103801} (\bibinfo {year} {2025})}\BibitemShut {NoStop}%
\bibitem [{\citenamefont {Bhat}\ \emph {et~al.}(2024)\citenamefont {Bhat}, \citenamefont {Das}, \citenamefont {Ravishankar},\ and\ \citenamefont {Paradkar}}]{KNmine}%
  \BibitemOpen
  \bibfield  {author} {\bibinfo {author} {\bibfnamefont {S.}~\bibnamefont {Bhat}}, \bibinfo {author} {\bibfnamefont {A.}~\bibnamefont {Das}}, \bibinfo {author} {\bibfnamefont {V.}~\bibnamefont {Ravishankar}},\ and\ \bibinfo {author} {\bibfnamefont {B.}~\bibnamefont {Paradkar}},\ }\bibfield  {title} {\bibinfo {title} {Novel instabilities in counter-streaming nonabelian fluids},\ }\href {https://doi.org/https://doi.org/10.1016/j.fpp.2024.100056} {\bibfield  {journal} {\bibinfo  {journal} {Fundamental Plasma Physics}\ }\textbf {\bibinfo {volume} {11}},\ \bibinfo {pages} {100056} (\bibinfo {year} {2024})}\BibitemShut {NoStop}%
\bibitem [{Note1()}]{Note1}%
  \BibitemOpen
  \bibinfo {note} {For the background review of Yang-Mills dynamics, one can refer to the following standard treatments \cite {WongClassicalYM-Isospin,Boozer,YMSpinTAN,KNmine}}\BibitemShut {NoStop}%
\bibitem [{\citenamefont {Wong}(1970)}]{WongClassicalYM-Isospin}%
  \BibitemOpen
  \bibfield  {author} {\bibinfo {author} {\bibfnamefont {S.~K.}\ \bibnamefont {Wong}},\ }\bibfield  {title} {\bibinfo {title} {Field and particle equations for the classical yang-mills field and particles with isotopic spin},\ }\href {https://doi.org/10.1007/BF02892134} {\bibfield  {journal} {\bibinfo  {journal} {Il Nuovo Cimento A (1965-1970)}\ }\textbf {\bibinfo {volume} {65}},\ \bibinfo {pages} {689} (\bibinfo {year} {1970})}\BibitemShut {NoStop}%
\bibitem [{\citenamefont {Wu}\ and\ \citenamefont {Yang}(1975)}]{WuYang}%
  \BibitemOpen
  \bibfield  {author} {\bibinfo {author} {\bibfnamefont {T.~T.}\ \bibnamefont {Wu}}\ and\ \bibinfo {author} {\bibfnamefont {C.~N.}\ \bibnamefont {Yang}},\ }\bibfield  {title} {\bibinfo {title} {Some remarks about unquantized non-abelian gauge fields},\ }\href {https://doi.org/10.1103/PhysRevD.12.3843} {\bibfield  {journal} {\bibinfo  {journal} {Phys. Rev. D}\ }\textbf {\bibinfo {volume} {12}},\ \bibinfo {pages} {3843} (\bibinfo {year} {1975})}\BibitemShut {NoStop}%
\bibitem [{\citenamefont {Bhatia}(1997)}]{bhatia97}%
  \BibitemOpen
  \bibfield  {author} {\bibinfo {author} {\bibfnamefont {R.}~\bibnamefont {Bhatia}},\ }\href@noop {} {\emph {\bibinfo {title} {Matrix Analysis}}},\ Vol.\ \bibinfo {volume} {169}\ (\bibinfo  {publisher} {Springer},\ \bibinfo {year} {1997})\BibitemShut {NoStop}%
\bibitem [{\citenamefont {Whittaker}\ and\ \citenamefont {Watson}(2020)}]{whittaker2020course}%
  \BibitemOpen
  \bibfield  {author} {\bibinfo {author} {\bibfnamefont {E.}~\bibnamefont {Whittaker}}\ and\ \bibinfo {author} {\bibfnamefont {G.}~\bibnamefont {Watson}},\ }\href {https://books.google.co.in/books?id=DK7qDwAAQBAJ} {\emph {\bibinfo {title} {A Course of Modern Analysis}}},\ Dover Books on Mathematics\ (\bibinfo  {publisher} {Dover Publications},\ \bibinfo {year} {2020})\BibitemShut {NoStop}%
\bibitem [{\citenamefont {Zohar}\ \emph {et~al.}(2013)\citenamefont {Zohar}, \citenamefont {Cirac},\ and\ \citenamefont {Reznik}}]{zohar2013quantum}%
  \BibitemOpen
  \bibfield  {author} {\bibinfo {author} {\bibfnamefont {E.}~\bibnamefont {Zohar}}, \bibinfo {author} {\bibfnamefont {J.~I.}\ \bibnamefont {Cirac}},\ and\ \bibinfo {author} {\bibfnamefont {B.}~\bibnamefont {Reznik}},\ }\bibfield  {title} {\bibinfo {title} {Quantum simulations of gauge theories with ultracold atoms: Local gauge invariance from angular-momentum conservation},\ }\href@noop {} {\bibfield  {journal} {\bibinfo  {journal} {Physical Review A—Atomic, Molecular, and Optical Physics}\ }\textbf {\bibinfo {volume} {88}},\ \bibinfo {pages} {023617} (\bibinfo {year} {2013})}\BibitemShut {NoStop}%
\bibitem [{\citenamefont {Halimeh}\ \emph {et~al.}(2022)\citenamefont {Halimeh}, \citenamefont {McCulloch}, \citenamefont {Yang},\ and\ \citenamefont {Hauke}}]{PRXQuantum.QCD}%
  \BibitemOpen
  \bibfield  {author} {\bibinfo {author} {\bibfnamefont {J.~C.}\ \bibnamefont {Halimeh}}, \bibinfo {author} {\bibfnamefont {I.~P.}\ \bibnamefont {McCulloch}}, \bibinfo {author} {\bibfnamefont {B.}~\bibnamefont {Yang}},\ and\ \bibinfo {author} {\bibfnamefont {P.}~\bibnamefont {Hauke}},\ }\bibfield  {title} {\bibinfo {title} {Tuning the topological $\ensuremath{\theta}$-angle in cold-atom quantum simulators of gauge theories},\ }\href {https://doi.org/10.1103/PRXQuantum.3.040316} {\bibfield  {journal} {\bibinfo  {journal} {PRX Quantum}\ }\textbf {\bibinfo {volume} {3}},\ \bibinfo {pages} {040316} (\bibinfo {year} {2022})}\BibitemShut {NoStop}%
\bibitem [{\citenamefont {Boozer}(2011)}]{Boozer}%
  \BibitemOpen
  \bibfield  {author} {\bibinfo {author} {\bibfnamefont {A.~D.}\ \bibnamefont {Boozer}},\ }\bibfield  {title} {\bibinfo {title} {Classical yang-mills theory},\ }\href {https://doi.org/10.1119/1.3606478} {\bibfield  {journal} {\bibinfo  {journal} {American Journal of Physics}\ }\textbf {\bibinfo {volume} {79}},\ \bibinfo {pages} {925} (\bibinfo {year} {2011})},\ \Eprint {https://arxiv.org/abs/https://doi.org/10.1119/1.3606478} {https://doi.org/10.1119/1.3606478} \BibitemShut {NoStop}%
\end{thebibliography}

%

\appendix
\section{\label{app:config3}Particle Motion in Field Configuration \RNum{3}}
As mentioned in the main text, the most general combination of maximally non-Abelian electric and magnetic fields is realised when all components of the vector potential and scalar potentials are present. Consider the vector potential that results in the three-component color magnetic field \eqref{eq:3comp-a}; to this, a general scalar potential is introduced. The field configuration is as follows
\begin{eqnarray}
    \label{eq:pot-config1}
    \vec{A}_x &=& A_x \hat{e}_1, \qquad
    \vec{A}_y = A_y \hat{e}_2, \qquad
    \vec{A}_z = A_z \hat{e}_3, \nonumber \\
    \vec{\phi} &=& \phi_1 \hat{e}_1 + \phi_2 \hat{e}_2 + \phi_3 \hat{e}_3 
\end{eqnarray}
The associated color fields are
\begin{eqnarray}
    \label{eq:field-config1}
    \vec{B}_x &=& B_x \hat{e}_1, \qquad
    \vec{B}_y = B_y \hat{e}_2, \qquad
    \vec{B}_z = B_z \hat{e}_3, \nonumber \\
    \vec{E}_x &=& E_{x,2} \hat{e}_2 - E_{x,3} \hat{e}_3, \qquad
    \vec{E}_y = E_{y,3} \hat{e}_3 - E_{y,1} \hat{e}_1, \nonumber \\
    \vec{E}_z &=& E_{z,1} \hat{e}_1 - E_{z,2} \hat{e}_2 
\end{eqnarray}

The resulting color electric and magnetic field will have all the spatial and color components and satisfy the bi-orthogonal property discussed in the main text. Here, we introduce another index to $\alpha_i$ to denote the color component, the notation to be followed in the appendix is $\alpha_{i, a} = \frac{E_{i, a}}{B}$, where $i$ denotes the spatial component and $a$ denotes the color component. A similar notation is used to denote the color electric field components in \eqref{eq:field-config1}.

Field configurations of this type are relevant primarily in high-energy contexts. In particular, they are expected to arise during the early-time evolution of a quark–gluon plasma, where strong chromo-electric and chromo-magnetic fields generated by the maximally non-Abelian term can be approximately homogeneous over short time and length scales. In such regimes, the dynamics of individual color charges are governed by the dynamics discussed here. Although direct experimental access to single particle trajectories in these systems is not possible, the classical test particle dynamics studied here provides useful intuition for how non-Abelian gauge interactions influence particle motion, charge evolution, and drift even in uniform backgrounds.

At the same time, this field configuration could be engineered in laboratory systems such as ultracold atoms and photonic systems. Since, in these systems, one could observe the trajectory of the particle and also the evolution of the color charge itself, the cold-atom and photonic systems could act as effective simulators of non-Abelian particle dynamics, bridging concepts from high-energy gauge theories and synthetic gauge field realisations within a classical framework.

Substituting the above fields into the dimensionless equations of motion, one obtains six coupled first-order differential equations. The components of canonical momentum and the total energy for this configuration are given by
\begin{eqnarray}
    \Pi_x &=& p_x + a_x \zeta_1, \qquad
    \Pi_y = p_y + a_y \zeta_2, \nonumber \\
    \Pi_z &=& p_z + a_z \zeta_3, \qquad
    \varepsilon = \gamma + \vec{\zeta} \cdot \vec{\varphi} .
\end{eqnarray}
These constants of motion impose bounds on the dynamical variables. Using $|\zeta_i (\tau)|\leq 1$, one finds

\begin{eqnarray}
    (\Pi_i - a_i) &\leq& p_i \leq (\Pi_i + a_i), \nonumber \\ 
    \left(\varepsilon - \sum_i \varphi_i \right) &\leq& \sqrt{1 + p^2}
    \leq \left(\varepsilon + \sum_i \varphi_i\right).
\end{eqnarray}

Employing these constants of motion, the full set of equations reduces to a closed system of three coupled equations for the particle momentum, with the evolution of the color charge $\vec{\zeta}(\tau)$ following that of the momentum. The resulting reduced equations are
\begin{widetext}
\begin{subequations}
    \label{eq:simp-config1}
    \begin{eqnarray}
        \frac{d p_x}{d \tau} &=& a_x(\varepsilon - \gamma) - \varphi_1(\Pi_x - p_x)
        + \frac{1}{\gamma}\left(p_y \mathcal{B}_z - p_z \mathcal{B}_y
        - p_y p_z\left(\frac{b_z}{a_z} - \frac{b_y}{a_y}\right)\right), \\
        \frac{d p_y}{d \tau} &=& a_y(\varepsilon - \gamma) - \varphi_2(\Pi_y - p_y)
        + \frac{1}{\gamma}\left(p_z \mathcal{B}_x - p_x \mathcal{B}_z
        - p_z p_x\left(\frac{b_x}{a_x} - \frac{b_z}{a_z}\right)\right), \\
        \frac{d p_z}{d \tau} &=& a_z(\varepsilon - \gamma) - \varphi_3(\Pi_z - p_z)
        + \frac{1}{\gamma}\left(p_x \mathcal{B}_y - p_y \mathcal{B}_x
        - p_x p_y\left(\frac{b_y}{a_y} - \frac{b_x}{a_x}\right)\right).
    \end{eqnarray}
\end{subequations}
\end{widetext}

\begin{figure*}[h!]
    \centering
    \begin{subfigure}{0.4\textwidth}
        \centering
        \includegraphics[width=\linewidth]{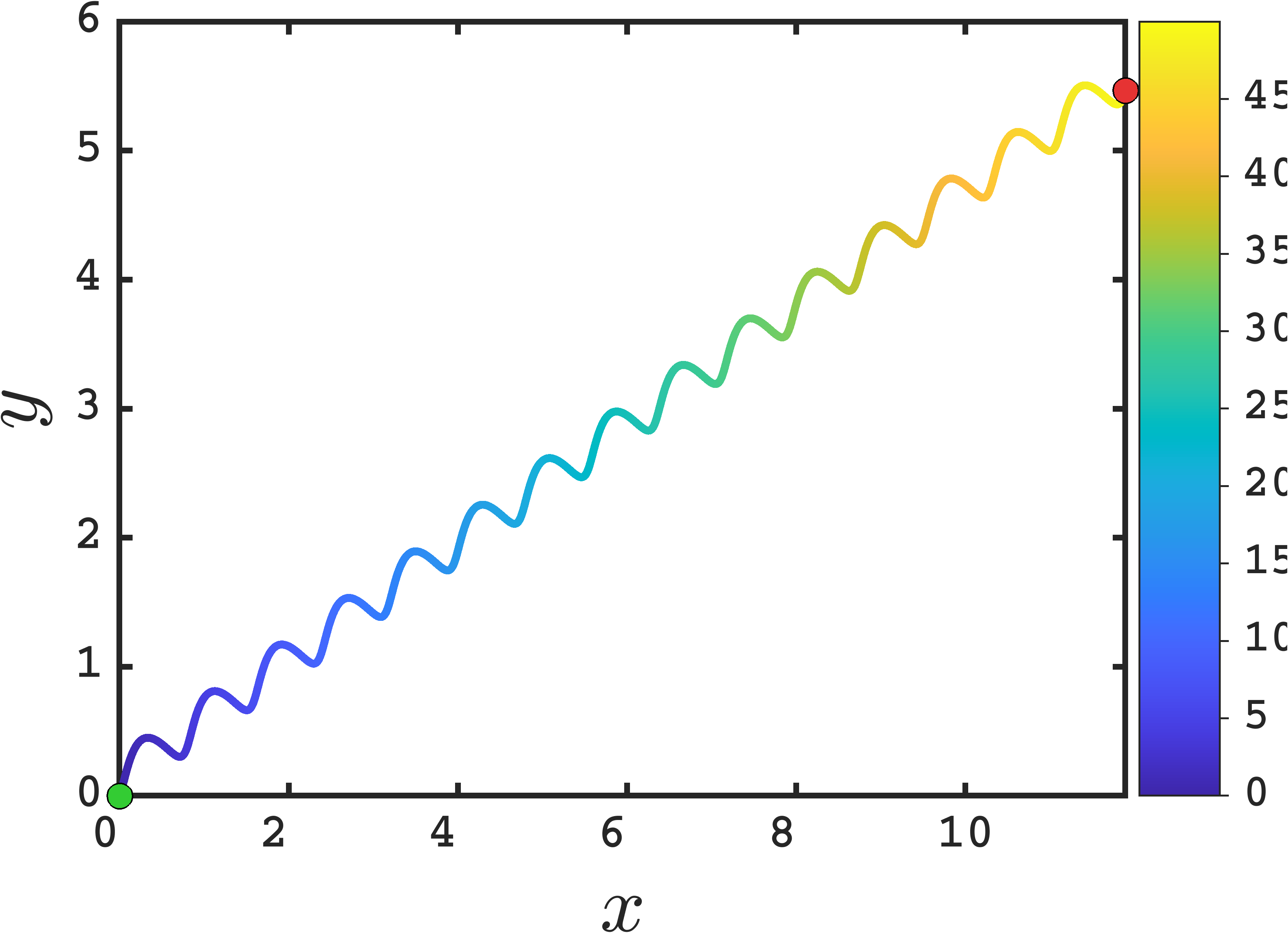}
        \caption{$x$–$y$ plane}
        \label{fig:c1-a-xy}
    \end{subfigure}
    \hfill
    \begin{subfigure}{0.4\textwidth}
        \centering
        \includegraphics[width=\linewidth]{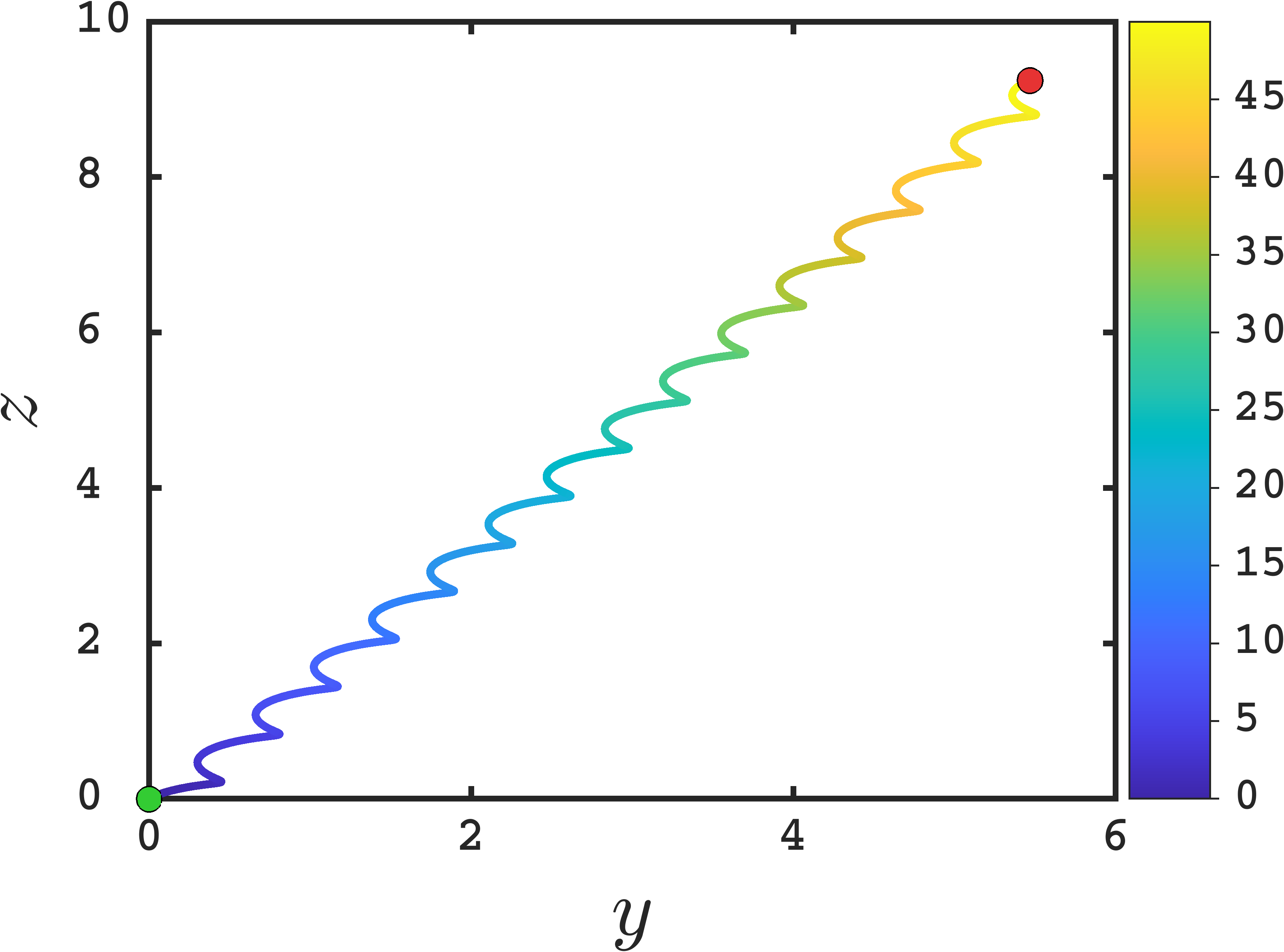}
        \caption{$y$–$z$ plane}
        \label{fig:c1-a-yz}
    \end{subfigure}
    \medskip
    \begin{subfigure}{0.4\textwidth}
        \centering
        \includegraphics[width=\linewidth]{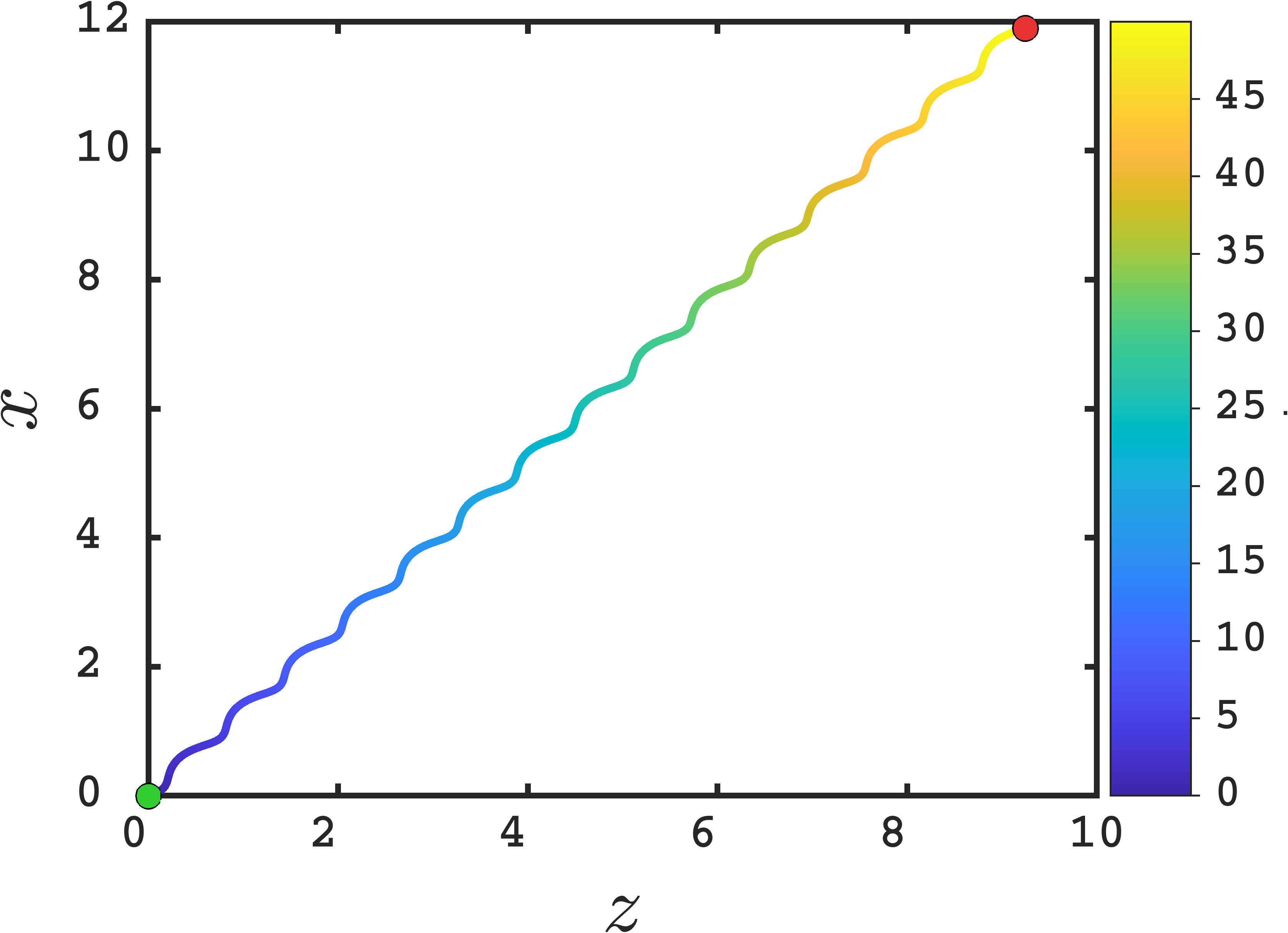}
        \caption{$z$–$x$ plane}
        \label{fig:c1-a-zx}
    \end{subfigure}
    \hfill
    \begin{subfigure}{0.4\textwidth}
        \centering
        \includegraphics[width=\linewidth]{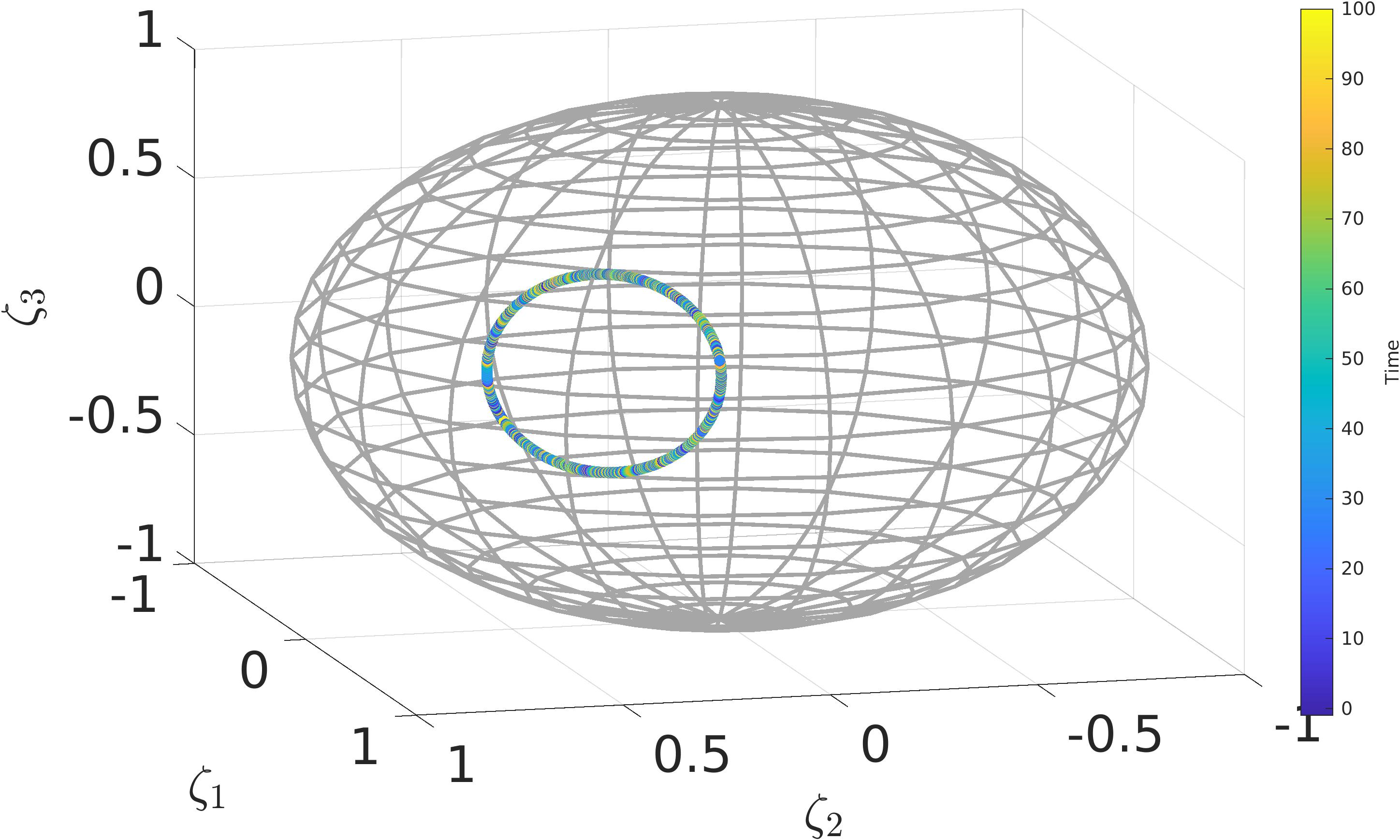}
        \caption{Color-charge evolution}
        \label{fig:c1-a-color}
    \end{subfigure}
    \caption{Particle dynamics in field configuration \RNum{3} for $\alpha_{x,2}=\alpha_{z,1}=\alpha_{y,3}=0.5$. Once one chooses these $\alpha_{i,a}$ the remaining ones are fixed. Parameters: $b_x=0.52$, $b_y=0.25$, $b_z=0.8$,$\Pi_x=0.55$, $\Pi_y=1.09$, $\Pi_z=0.55$, and $\varepsilon=2.55$.The color map represents time.}
    \label{fig:config3-alpha05}
\end{figure*}

The equations~\eqref{eq:simp-config1} are well suited for numerical integration. In contrast to the pure magnetic field case, no simple analogue of the $E\times B$ drift emerges. Although the particle generally exhibits a net drift, its direction and magnitude are not determined solely by the ratio of electric and magnetic field strengths, as in electrodynamics. This qualitative difference arises from the dynamical evolution of the color charge and its direct coupling to the gauge potentials. A representative particle trajectory and the corresponding color charge evolution are shown in Fig.~\ref{fig:config3-alpha05}.

The analysis presented here completes the classification of classical test particle dynamics in constant non-Abelian field configurations of pure YM origin. These results provide a natural starting point for future quantum mechanical studies and place non-Abelian test particle dynamics on a footing comparable to that of electrodynamics, while highlighting the fundamentally richer structure introduced by internal gauge degrees of freedom.

\end{document}